\documentclass[10pt, aps,pra,twocolumn,nofootinbib]{revtex4-2}

\usepackage{amsmath,amssymb,mathtools,bm}
\usepackage{graphicx}
\usepackage{booktabs}
\usepackage{microtype}
\usepackage{xcolor}
\usepackage{siunitx}
\usepackage{hyperref}
\usepackage{placeins}
\usepackage{tikz}
\usetikzlibrary{arrows.meta,positioning,calc,fit,shapes.geometric}

\hypersetup{
  colorlinks=true,
  citecolor=blue,
  linkcolor=blue,
  urlcolor=blue,
  pdftitle={Quantum Decision Theory for Displacement Sensing with Finite-Energy GKP Codes: Bayesian and Neyman--Pearson Limits under Loss and Amplification},
  pdfsubject={Equal-squeezing Bayesian and Neyman--Pearson comparison of finite-energy GKP, coherent, squeezed, and twin-beam probes}
}
\graphicspath{{figures/}}
\newcommand{\Tr}{\operatorname{Tr}}

\newcommand{\id}{\mathbb{I}}
\newcommand{\cE}{\mathcal{E}}

\newcommand{\cU}{\mathcal{U}}
\newcommand{\cL}{\mathcal{L}}
\newcommand{\cA}{\mathcal{A}}
\newcommand{\cG}{\mathcal{G}}
\newcommand{\cN}{\mathcal{N}}
\newcommand{\cW}{\mathcal{W}}
\newcommand{\cV}{\mathcal{V}}

\newcommand{\Pe}{P_{\rm e}}
\newcommand{\PD}{P_{\rm D}}
\newcommand{\ketbra}[2]{|#1\rangle\!\langle #2|}
\newcommand{\pos}[1]{\left[#1\right]_{+}}
\newcommand{\dd}{\mathrm{d}}
\newcommand{\norm}[1]{\left\lVert#1\right\rVert}

\begin{document}

\title{Quantum Decision Theory for Displacement Detection with Finite-Energy GKP States}

\author{Seid Koudia}
\email{seid.koudia@uni.lu}
\affiliation{Interdisciplinary Centre for Security, Reliability and Trust (SnT), University of Luxembourg, Luxembourg}
\author{Symeon Chatzinotas}
\affiliation{Interdisciplinary Centre for Security, Reliability and Trust (SnT), University of Luxembourg, Luxembourg}

\date{6 August 2026}

\begin{abstract}
We develop a quantum-decision-theoretic framework for detecting phase-space displacements with finite-energy, $d$-level Gottesman--Kitaev--Preskill (GKP) probes. For single-mode and entanglement-assisted architectures, we derive the Bayesian minimum-error probability, the optimal Neyman--Pearson receiver-operating characteristic, and the corresponding minimum detectable displacement. Finite-energy effects are treated through exact theta-series displacement kernels, while pure loss followed by quantum-limited amplification is mapped to an effective Gaussian random-displacement channel. Entanglement removes preparation-dependent blind directions and preserves both logical displacement labels, although it does not surpass the pointwise optimized single-mode strategy in the noiseless pure-state setting. We benchmark the resulting protocols against coherent-state, direction-matched squeezed-vacuum, and twin-beam schemes at equal nominal squeezing. Numerical results identify finite-squeezing and lossy regimes in which GKP probes achieve both a lower Bayesian error and a smaller minimum detectable perturbation than the selected Gaussian receivers.
\end{abstract}

\maketitle

% ================================================================
\section{Introduction}
\label{sec:intro}

Quantum process estimation and discrimination provide operational frameworks for identifying physical transformations through their action on quantum probes. Whereas estimation seeks to infer continuous parameters of an unknown channel, discrimination asks which transformation, among a prescribed set of alternatives, has acted on the system. The latter is naturally formulated as quantum hypothesis testing: a probe is prepared, subjected to the unknown process, and measured to decide between competing hypotheses. Two complementary strategies are particularly relevant \cite{Maffeis2019}. A Bayesian receiver minimizes the average decision error for assigned prior probabilities \cite{Koudia2019CausalDisc}, while a Neyman--Pearson receiver maximizes the detection probability subject to a fixed false-alarm constraint \cite{Helstrom1976,Watrous2018}. These approaches have been applied to unitary-process discrimination, quantum communication, interferometry, target detection, and minimum-detectable-perturbation problems \cite{Acin2001,Dariano2002,Maffeis2019}.

Bosonic systems provide a natural setting for extending this decision-theoretic viewpoint to continuous-variable transformations. Here we consider binary discrimination between the absence and presence of a phase-space displacement acting on a single bosonic mode. Rather than estimating the displacement vector with minimum quadratic error, the objective is to determine whether a displacement occurred and, within a Neyman--Pearson formulation, the smallest perturbation that can be reliably detected. The displacement amplitude and direction may therefore appear as unknown or nuisance parameters. Standard Gaussian probes provide natural benchmarks: coherent states define a classical optical reference, squeezed states enhance sensitivity along a selected quadrature, and two-mode squeezed-vacuum states exploit signal--idler correlations and joint detection \cite{Weedbrook2012,Lloyd2008QI,Tan2008QI}.

Structured non-Gaussian probes can exhibit responses that are inaccessible to Gaussian states \cite{Koudia2021NGaussCausal}. Fock states, finite photon-number superpositions, photon-subtracted or photon-added states, and related non-Gaussian resources have consequently been investigated for loss estimation, phase sensing, displacement detection, quantum illumination, and target detection \cite{Adesso2009Loss,Wolf2019FockSensing,Hanamura2021NonGaussianDisplacement,Samantaray2020PhotonSubtracted,DiMario2020NonGaussianMeasurements,Zhang2024NonGaussianQI,Xu2022NonGaussianMetrology}. These results motivate probes whose non-Gaussian phase-space structure is adapted specifically to displacement transformations.

Gottesman--Kitaev--Preskill (GKP) states are particularly suited to this task because they are defined by periodic phase-space translations. A GKP code is an oscillator stabilizer code whose code space is fixed by commuting displacement operators, so a physical displacement appears directly through changes in modular stabilizer observables \cite{GKP2001,Koudia2025MIMO}. This property was exploited for simultaneous estimation of conjugate displacement components \cite{Duivenvoorden2017}, and subsequent work has developed grid-state and modular-observable approaches to multiparameter sensing and displacement estimation \cite{Valahu2025,Labarca2026}. For process discrimination, the lattice structure provides an additional discrete layer: a displacement can be decomposed into a modular syndrome within a fundamental cell, a logical Weyl class associated with the quotient of the symplectic dual lattice by the stabilizer lattice, and a stabilizer translation acting trivially on the encoded information \cite{GKP2001,Conrad2022,ConradThesis2024,Shaw2024, koudia2026measurement}. This makes both the lattice geometry and the encoded dimension relevant sensing parameters, since they control the separation of logical displacement classes and their stabilizer aliases.

Realistic GKP probes have finite energy: their peaks have finite width, the grid is modulated by an envelope, and approximate logical codewords need not be exactly orthogonal. Consequently, finite squeezing modifies stabilizer periodicity, logical-state overlaps, and leakage outside the approximate code space \cite{Glancy2006,Matsuura2020,Tzitrin2020,Royer2020}. These effects must be retained when evaluating Bayesian error probabilities, Neyman--Pearson receiver-operating characteristics, and minimum detectable perturbations. We also consider photon loss, which attenuates the probe and displacement signal and acts as a non-Pauli bosonic channel on finite-energy grid states. When followed by quantum-limited amplification with the appropriate gain, the resulting unit-gain channel can be represented as Gaussian random-displacement noise, placing loss compensation in the same phase-space framework as GKP decoding \cite{Weedbrook2012,Fukui2021LossAmp,Hastrup2023Loss,Harris2025Loss,Zheng2025LossAmp}.

\begin{figure*}[t]
    \centering
    \includegraphics[width=0.98\textwidth]{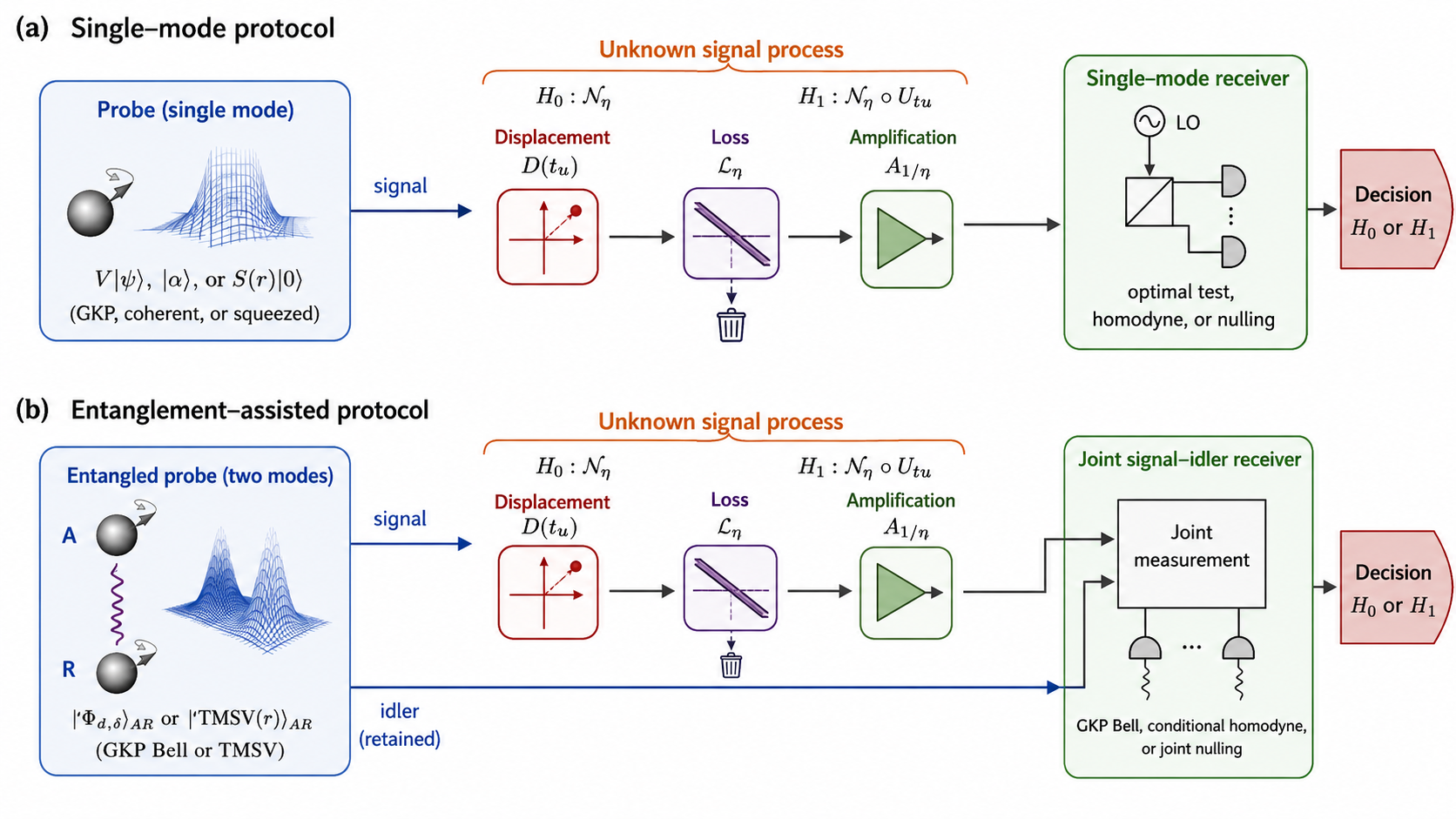}
    \caption{Binary displacement-discrimination schemes. (a) Single-mode protocol with a receiver acting only on the signal mode. (b) Entanglement-assisted protocol in which the signal traverses the unknown process, the idler is retained locally, and both modes are measured jointly.}
    \label{fig:protocol}
\end{figure*}

We study the two GKP sensing architectures shown schematically in Fig.~\ref{fig:protocol}. In the single-mode scheme, a finite-energy GKP state is subjected to the unknown displacement and measured locally. Its discrimination performance can depend strongly on the logical preparation and may exhibit preparation-dependent blind directions. In the entanglement-assisted scheme, the signal mode is maximally entangled with a retained GKP idler and only the signal undergoes the unknown process. Joint signal--idler measurement maps logical displacement classes onto logical-Bell sectors, thereby removing the dependence on an arbitrary single-mode logical input. The entangled scheme is nevertheless compared against the pointwise optimized single-mode strategy, particularly in the noiseless deterministic setting.

The GKP protocols are benchmarked against coherent, direction-matched squeezed-vacuum, and twin-beam probes. Quadrature-based and inverse-preparation vacuum-or-not receivers are considered where appropriate. Comparisons are made at equal nominal squeezing, with signal and total photon energies reported separately, allowing the roles of Gaussian squeezing, finite-energy grid structure, lattice geometry, and logical dimension to be distinguished.

Within this framework, we derive Bayesian error probabilities and Neyman--Pearson receiver-operating characteristics for single-mode and entanglement-assisted GKP probes, determine their minimum detectable displacements, and characterize the influence of finite squeezing, logical dimension, probe preparation, and loss. For the loss-compensated channel, the ideal GKP problem reduces to syndrome-resolved wrapped likelihoods, whereas finite-energy mixed outputs are treated directly at the oscillator level. The resulting analysis identifies when the modular and logical structure of GKP probes improves displacement detectability relative to Gaussian benchmarks.

The remainder of the paper is organized as follows. Section~\ref{sec:prelim} reviews Bayesian and Neyman--Pearson quantum discrimination and introduces the bosonic, Gaussian, and GKP tools used throughout the analysis. Section~\ref{sec:bayes} develops Bayesian process detection for single-mode and entanglement-assisted GKP probes, including Gaussian benchmarks and loss followed by amplification. Section~\ref{sec:np} derives the corresponding receiver-operating characteristics and minimum detectable perturbations. Section~\ref{sec:conclusion} discusses the operational
implications, limitations, and possible extensions of the framework.

% ================================================================
\section{Mathematical preliminaries}
\label{sec:prelim}

\subsection{Bayesian quantum process discrimination}

Let $\rho_0$ and $\rho_1$ be the two possible output states, and let the POVM element $Q$ denote the decision $H_1$. With priors $z_0+z_1=1$, the average error probability is
\begin{align*}
 q_{\rm e}(Q)
 &=z_0\Tr(Q\rho_0)+z_1\Tr[(\id-Q)\rho_1]\\
 &=z_1-\Tr(Q\Lambda_z),
\end{align*}
where
\begin{equation*}
 \Lambda_z=z_1\rho_1-z_0\rho_0
\end{equation*}
is the Bayesian characteristic operator. The minimum is attained when $Q$ projects onto the positive spectral subspace of $\Lambda_z$, yielding
\begin{equation}
 \Pe^\star=\frac12\left(1-\norm{\Lambda_z}_1\right).
 \label{eq:helstrom}
\end{equation}

For two pure output states, write $\rho_h=|\Psi_h\rangle\langle\Psi_h|$ and define
\begin{equation}
 \kappa=\langle\Psi_0|\Psi_1\rangle,
 \qquad
 s=\sqrt{1-|\kappa|^2}.
 \label{eq:pure-overlap-prelim}
\end{equation}
Choosing $|\Psi_0^\perp\rangle$ such that
\begin{equation*}
 |\Psi_1\rangle
 =\kappa|\Psi_0\rangle+s|\Psi_0^\perp\rangle,
\end{equation*}
the characteristic operator restricted to the support of the two hypotheses is
\begin{equation}
 \Lambda_z=
 \begin{pmatrix}
 z_1|\kappa|^2-z_0 & z_1\kappa s\\
 z_1\kappa^*s & z_1s^2
 \end{pmatrix}.
 \label{eq:pure-helstrom-matrix}
\end{equation}
Its trace and determinant are
\begin{equation*}
 \Tr\Lambda_z=z_1-z_0,
 \qquad
 \det\Lambda_z=-z_0z_1(1-|\kappa|^2),
\end{equation*}
and its eigenvalues are
\begin{equation*}
 \lambda_\pm=\frac12\left[
 z_1-z_0
 \pm\sqrt{1-4z_0z_1|\kappa|^2}
 \right].
\end{equation*}
Equation~\eqref{eq:helstrom} therefore becomes
\begin{equation}
 \Pe^\star=\frac12\left[
 1-\sqrt{1-4z_0z_1|\kappa|^2}
 \right].
 \label{eq:pure-bayes}
\end{equation}
The noiseless GKP results derived below follow by evaluating the physical overlap $\kappa$ and substituting it into Eq.~\eqref{eq:pure-bayes}. If the displacement parameter has a nuisance prior $\pi(\bm\xi)$, the alternative state is replaced by
\begin{equation*}
 \rho_1\longrightarrow
 \int\pi(\bm\xi)\rho_1(\bm\xi)\dd^2\bm\xi,
\end{equation*}
while Eq.~\eqref{eq:helstrom} remains valid.

\subsection{Neyman--Pearson quantum discrimination}

The Neyman--Pearson strategy fixes an admissible false-alarm probability $\alpha$ and maximizes the probability of detecting $H_1$. With $Q$ again denoting the decision $H_1$,
\begin{equation*}
 p_{10}=\Tr(Q\rho_0),
 \qquad
 p_{11}=\Tr(Q\rho_1).
\end{equation*}
Introducing a Lagrange multiplier $\gamma\geq0$ gives the characteristic operator
\begin{equation*}
 \Gamma_\gamma=\rho_1-\gamma\rho_0.
\end{equation*}
For fixed $\gamma$, the optimal test projects onto the positive spectral subspace of $\Gamma_\gamma$, with possible randomization on its kernel. Equivalently,
\begin{equation*}
 \PD^\star(\alpha)=
 \min_{\gamma\geq0}\left\{
 \gamma\alpha+\Tr\pos{\rho_1-\gamma\rho_0}
 \right\}.
\end{equation*}

For the pure states introduced in Eq.~\eqref{eq:pure-overlap-prelim},
\begin{equation*}
 \Gamma_\gamma=
 \begin{pmatrix}
 |\kappa|^2-\gamma & \kappa s\\
 \kappa^*s & s^2
 \end{pmatrix},
\end{equation*}
with eigenvalues
\begin{align}
 g_\pm(\gamma)
 &=\frac12\left[1-\gamma\pm\Delta_\gamma\right],
 \label{eq:pure-np-eigs}\\
 \Delta_\gamma
 &=\sqrt{(1+\gamma)^2-4\gamma|\kappa|^2}.
 \notag
\end{align}
For $0<|\kappa|<1$, the determinant is negative, so the optimal test is rank one. Using the spectral projector
\begin{equation*}
 Q_\gamma=\frac{\Gamma_\gamma-g_-\id}{\Delta_\gamma},
\end{equation*}
one obtains the parametric characteristic curve
\begin{align*}
 p_{10}(\gamma)
 &=\frac12\left[
 1-\frac{1+\gamma-2|\kappa|^2}{\Delta_\gamma}
 \right],
 \\
 p_{11}(\gamma)
 &=\frac12\left[
 1+\frac{1+\gamma-2\gamma|\kappa|^2}{\Delta_\gamma}
 \right].
\end{align*}

The normalized positive eigenvector can be written as
\begin{equation}
 |g_+\rangle=
 \frac{
 \kappa s|\Psi_0\rangle+
 [g_+-(|\kappa|^2-\gamma)]|\Psi_0^\perp\rangle
 }{
 \sqrt{
 |\kappa|^2s^2+
 [g_+-(|\kappa|^2-\gamma)]^2
 }
 },
 \label{eq:pure-np-eigenvector}
\end{equation}
so that $Q_\gamma=|g_+\rangle\langle g_+|$ defines an explicit optimal receiver. Alternatively, setting
\begin{equation*}
 p_{10}=\sin^2\theta,
 \qquad
 1-|\kappa|^2=\sin^2\varphi,
\end{equation*}
with $0\leq\theta+\varphi\leq\pi/2$, gives
\begin{equation*}
 p_{11}=\sin^2(\theta+\varphi).
\end{equation*}

Eliminating $\gamma$ yields the optimal receiver-operating characteristic on the nontrivial branch:
\begin{equation}
 p_{11}^\star(p_{10})=
 \left[
 \sqrt{p_{10}}|\kappa|
 +\sqrt{1-p_{10}}\sqrt{1-|\kappa|^2}
 \right]^2.
 \label{eq:pure-roc-prelim}
\end{equation}
This expression holds for
$0\leq p_{10}\leq|\kappa|^2$; for
$|\kappa|^2<p_{10}\leq1$, the optimum is
$p_{11}^\star=1$. Thus, both the Bayesian error in Eq.~\eqref{eq:pure-bayes} and the Neyman--Pearson characteristic in Eq.~\eqref{eq:pure-roc-prelim} are determined by the physical overlap $\kappa$.

\subsection{Displacements, loss, and amplification}

We use the quadrature vector
\begin{equation*}
 \hat{\bm R}=(\hat q,\hat p)^T,
 \qquad
 [\hat q,\hat p]=i,
\end{equation*}
and define
\begin{equation*}
 D(\bm\xi)=
 \exp\!\left(i\hat{\bm R}^{T}\Omega\bm\xi\right),
 \qquad
 \Omega=
 \begin{pmatrix}
 0&1\\
 -1&0
 \end{pmatrix}.
\end{equation*}
The Weyl product is
\begin{equation*}
 D(\bm\xi)D(\bm\zeta)=
 e^{-i\bm\xi^T\Omega\bm\zeta/2}
 D(\bm\xi+\bm\zeta).
\end{equation*}
For the symmetric characteristic function
\begin{equation*}
 \chi_\rho(\bm k)=\Tr[\rho D(\bm k)],
\end{equation*}
pure loss of transmissivity $\eta$ and quantum-limited amplification of gain $G$ act as
\begin{align}
 \chi_{\cL_\eta(\rho)}(\bm k)
 &=\chi_\rho(\sqrt\eta\,\bm k)
 e^{-(1-\eta)\norm{\bm k}^2/4},
 \label{eq:loss-chi}\\
 \chi_{\cA_G(\rho)}(\bm k)
 &=\chi_\rho(\sqrt G\,\bm k)
 e^{-(G-1)\norm{\bm k}^2/4}.
 \label{eq:amp-chi}
\end{align}

Post-amplification with $G=1/\eta$ gives
\begin{equation*}
 \cA_{1/\eta}\circ\cL_\eta
 =\cG_{\sigma_\eta^2},
 \qquad
 \sigma_\eta^2=\frac{1-\eta}{\eta},
\end{equation*}
where
\begin{equation}
 \cG_{\sigma^2}(\rho)=
 \int_{\mathbb R^2}
 \frac{\dd^2\bm\nu}{2\pi\sigma^2}
 e^{-\norm{\bm\nu}^2/(2\sigma^2)}
 D(\bm\nu)\rho D^\dagger(\bm\nu).
 \label{eq:additive-channel}
\end{equation}
If amplification precedes loss, the corresponding unit-gain variance is
\begin{equation*}
 \sigma_{\rm pre}^2=1-\eta.
\end{equation*}
Direct loss without gain compensation is treated in Appendix~\ref{app:channels}; unlike the compensated channel, it includes the phase-space contraction
$\bm k\mapsto\sqrt\eta\,\bm k$.

\subsection{Selected Gaussian benchmark probes and equal-squeezing convention}
\label{subsec:gaussian-prelim}

We use three Gaussian benchmark probes. Quadrature covariance matrices are defined by
\begin{equation*}
 V_{jk}=
 \frac12\langle
 \{\Delta\hat R_j,\Delta\hat R_k\}
 \rangle,
 \qquad
 V_{\rm vac}=\frac12\id_2.
\end{equation*}
A coherent displacement common to both hypotheses can be undone at the receiver, so the coherent-state benchmark can be represented by the vacuum without changing the discrimination performance.

The direction-matched squeezed-vacuum probe is
\begin{equation*}
 |0;r,\varphi\rangle
 =S(re^{i2\varphi})|0\rangle,
 \qquad
 \bar n_{\rm G}=\sinh^2r,
\end{equation*}
where the squeezing angle $\varphi$ is chosen to maximize sensitivity along the known displacement direction.

The entangled Gaussian benchmark is the twin-beam state
\begin{equation*}
 |\mathrm{TWB}(r)\rangle
 =\sqrt{1-\lambda^2}
 \sum_{n=0}^{\infty}
 \lambda^n|n\rangle_A|n\rangle_R,
 \qquad
 \lambda=\tanh r.
\end{equation*}
Each mode contains
$\bar n_{\rm G}=\sinh^2r$ photons. Defining
\begin{equation*}
 c=\cosh2r,
 \qquad
 s=\sinh2r,
 \qquad
 Z=\operatorname{diag}(1,-1),
\end{equation*}
its covariance matrix is
\begin{equation*}
 V_{\rm TWB}=
 \frac12
 \begin{pmatrix}
 c\id_2&sZ\\
 sZ&c\id_2
 \end{pmatrix}.
\end{equation*}
Displacing only the signal mode gives
\begin{equation*}
 \langle\mathrm{TWB}|
 D_A(\bm\xi)
 |\mathrm{TWB}\rangle
 =
 \exp\!\left[
 -\frac{(2\bar n_{\rm G}+1)\norm{\bm\xi}^2}{4}
 \right].
\end{equation*}

The equal-squeezing convention is defined by
\begin{equation}
 v_s\equiv10^{-s_{\rm dB}/10},
 \qquad
 \tanh\beta=v_s=e^{-2r}.
 \label{eq:equal-squeezing-rule}
\end{equation}
The corresponding Gaussian signal energy is
\begin{equation*}
 \bar n_{\rm G}(s_{\rm dB})
 =\sinh^2r
 =\frac14\left(v_s+v_s^{-1}-2\right),
\end{equation*}
whereas the finite-energy GKP signal energy is obtained from the physical embedding:
\begin{equation*}
 \bar n_{\rm GKP}^{(d)}(s_{\rm dB})
 =\frac1d\Tr[V^\dagger\hat n V].
\end{equation*}
The total twin-beam energy is $2\bar n_{\rm G}$, while a physically encoded GKP Bell pair has twice the code-averaged single-mode energy. Since equal nominal squeezing does not imply equal photon number, both the signal and total energies are reported in the comparative sweeps.

\subsection{GKP stabilizer lattices and the $d$-level quotient}
\label{subsec:gkp-lattice}

A one-mode GKP stabilizer group is specified by a full-rank phase-space lattice
\begin{equation*}
 \Lambda=M\mathbb Z^2\subset\mathbb R^2
\end{equation*}
satisfying
\begin{equation*}
 \bm\lambda^T\Omega\bm\lambda'
 \in2\pi\mathbb Z,
 \qquad
 \bm\lambda,\bm\lambda'\in\Lambda.
\end{equation*}
The projective phases form part of the code specification. If
$\bm\lambda=M\bm n$ and
\begin{equation*}
 A=\frac{1}{2\pi}M^T\Omega M
 \in\mathbb Z^{2\times2},
\end{equation*}
an ordered generator convention gives
\begin{equation*}
 \phi_M(\bm\lambda)
 =\pi\bm n^T\underline A\bm n,
\end{equation*}
where $\underline A$ is the strictly lower-triangular part of $A$ \cite{Conrad2022,ConradThesis2024}. The stabilizer group is therefore
\begin{equation*}
 \mathsf S(\Lambda,\phi_M)=
 \left\{
 e^{i\phi_M(\bm\lambda)}D(\bm\lambda):
 \bm\lambda\in\Lambda
 \right\}.
\end{equation*}

The symplectic dual lattice is
\begin{equation*}
 \Lambda^\perp=
 \left\{
 \bm\zeta:
 \bm\zeta^T\Omega\bm\lambda
 \in2\pi\mathbb Z
 \quad
 \forall\bm\lambda\in\Lambda
 \right\}.
\end{equation*}
Logical displacements are the cosets of
$\Lambda^\perp/\Lambda$. For a one-mode code of type $d$,
\begin{equation*}
 |\Lambda^\perp/\Lambda|=d^2,
\end{equation*}
and the encoded space has dimension $d$.

Every one-mode type-$d$ lattice is symplectically equivalent to a square representative. In physical quadrature units, we take
\begin{equation*}
 \Lambda_d=\sqrt{2\pi d}\,\mathbb Z^2,
 \qquad
 \Lambda_d^\perp=\ell_d\mathbb Z^2,
 \qquad
 \ell_d=\sqrt{\frac{2\pi}{d}}.
\end{equation*}
The logical generators can be chosen as
\begin{equation*}
 \bar X=D(\ell_d,0),
 \qquad
 \bar Z=D(0,\ell_d),
\end{equation*}
with
\begin{equation*}
 \bar Z\bar X=
 e^{2\pi i/d}\bar X\bar Z,
\end{equation*}
up to the adopted phase convention. The ideal computational codewords are
\begin{equation}
 |j_L^{(d)}\rangle
 \propto
 \sum_{n\in\mathbb Z}
 |q=(j+dn)\ell_d\rangle,
 \qquad
 j=0,\ldots,d-1.
 \label{eq:ideal-combs}
\end{equation}

\subsection{Modular GKP sensor observables}
\label{subsec:modular-sensor}

For the square type-$d$ code, define the stabilizer period
\begin{equation*}
 L_d=d\ell_d=\sqrt{2\pi d}
\end{equation*}
and choose
\begin{equation*}
 S_q=e^{iL_d\hat q}=D(0,L_d),
 \qquad
 S_p=e^{-iL_d\hat p}=D(L_d,0).
\end{equation*}
These operators commute because $L_d^2=2\pi d$. For
\begin{equation*}
 D(u,v)=e^{-iu\hat p+iv\hat q},
\end{equation*}
conjugation gives
\begin{align}
 D^\dagger(u,v)S_qD(u,v)
 &=e^{iL_du}S_q,
 \label{eq:sq-phase-shift}\\
 D^\dagger(u,v)S_pD(u,v)
 &=e^{-iL_dv}S_p.
 \label{eq:sp-phase-shift}
\end{align}
The two displacement components are therefore encoded in commuting modular eigenphases.

For $d=1$, this construction reduces to the grid-state sensor of Ref.~\cite{Duivenvoorden2017}. For a type-$d$ code, the phases determine $u$ and $v$ modulo
\begin{equation*}
 \frac{2\pi}{L_d}=\ell_d,
\end{equation*}
while the residue class in
$\Lambda_d^\perp/\Lambda_d$ identifies the logical Weyl sector. The natural analog-syndrome cell is
\begin{equation*}
 \cV_d=[-\ell_d/2,\ell_d/2)^2.
\end{equation*}
An ideal receiver may therefore resolve the modular syndrome and, when a logical code is used, the corresponding logical coset.

For reference, an approximate grid sensor of the form considered in Ref.~\cite{Duivenvoorden2017} is
\begin{equation*}
 |\psi_\Delta^{\rm grid}\rangle
 \propto
 \sum_{m\in\mathbb Z}
 e^{-\pi\Delta^2m^2}
 \int\dd q\,
 e^{-[q-\sqrt{2\pi}m]^2/(2\Delta^2)}
 |q\rangle.
\end{equation*}
The finite-energy family introduced below extends this construction to $d$ logical residue classes and provides an orthonormal physical embedding.

\subsection{Finite energy, theta kernels, and canonical embedding}
\label{subsec:finite-gkp}

We regularize the ideal codewords with the Fock envelope
\begin{equation*}
 |w_{j,\beta}\rangle
 =e^{-\beta\hat n}|j_L^{(d)}\rangle,
 \qquad
 \beta>0.
\end{equation*}
Defining
\begin{equation*}
 q_{jm}=(j+dm)\ell_d,
 \qquad
 v_\beta=\tanh\beta,
 \qquad
 c_\beta=\operatorname{sech}\beta,
\end{equation*}
the Mehler kernel gives the normalizable wavefunction
\begin{equation}
\begin{split}
 \psi_{j,\beta}(q)
 &=\mathcal N_{j,\beta}
 \sum_{m\in\mathbb Z}
 e^{-v_\beta q_{jm}^{2}/2}\\
 &\quad\times
 \exp\!\left[
 -\frac{(q-c_\beta q_{jm})^2}{2v_\beta}
 \right].
\end{split}
 \label{eq:finite-wavefunction}
\end{equation}
Finite energy broadens the peaks, suppresses distant peaks, and contracts their centers. The isolated-peak probability variance is $v_\beta/2$, and the corresponding effective squeezing is
\begin{equation*}
 s_{\rm dB}=-10\log_{10}v_\beta.
\end{equation*}

Collect the raw finite-energy codewords as the columns of
\begin{equation*}
 W=
 \left(
 |w_{0,\beta}\rangle,\ldots,
 |w_{d-1,\beta}\rangle
 \right),
 \qquad
 G=W^\dagger W.
\end{equation*}
Canonical symmetric orthonormalization defines the physical isometry
\begin{equation*}
 V=WG^{-1/2},
 \qquad
 V^\dagger V=\id_d,
\end{equation*}
and the compressed displacement operator
\begin{equation*}
 K_\beta^{(d)}(\bm\xi)
 =V^\dagger D(\bm\xi)V
 =G^{-1/2}B_\beta(\bm\xi)G^{-1/2}.
\end{equation*}

For the square code, $B_\beta$ can be evaluated analytically without a Fock-space truncation. Writing $\bm\xi=(x,p)^T$ and inserting Eq.~\eqref{eq:finite-wavefunction} gives
\begin{align}
 [B_\beta(x,p)]_{jk}
 &=\mathcal N_{j,\beta}\mathcal N_{k,\beta}
 \sqrt{\pi v_\beta}\,
 e^{-v_\beta p^2/4}
 \notag\\[-1mm]
 &\quad\times
 \sum_{m,n\in\mathbb Z}
 e^{-\mathcal Q_{jk}^{mn}
 +i\mathcal P_{jk}^{mn}},
 \label{eq:explicit-B-double-sum}\\
 \mathcal Q_{jk}^{mn}
 &=\frac{v_\beta}{2}
 \left(q_{jm}^2+q_{kn}^2\right)
 \notag\\[-1mm]
 &\quad+
 \frac{
 [c_\beta(q_{jm}-q_{kn})-x]^2
 }{4v_\beta},
 \notag\\
 \mathcal P_{jk}^{mn}
 &=\frac{c_\beta p}{2}
 \left(q_{jm}+q_{kn}\right).
 \notag
\end{align}
Equation~\eqref{eq:explicit-B-double-sum} is a convergent shifted two-dimensional theta series, and
\begin{equation*}
 G=B_\beta(0,0).
\end{equation*}
An equivalent logical-coset representation for a general symplectic lattice is derived in Appendix~\ref{app:theta}.

The same matrix can be written as the cross-ambiguity kernel
\begin{equation*}
 [B_\beta(x,p)]_{jk}
 =
 \int_{\mathbb R}\dd q\,
 \psi_{j,\beta}^*(q)
 e^{ip(q-x/2)}
 \psi_{k,\beta}(q-x).
\end{equation*}
Because $D(\bm\xi)V$ need not lie in the image of $V$,
$K_\beta^{(d)}(\bm\xi)$ is generally a contraction rather than a unitary. It nevertheless retains the overlaps and pairwise kernels required for the discrimination analysis.

\subsection{Single-mode and two-GKP entangled probes}
\label{subsec:entangled-basis}

The orthonormal finite-energy GKP logical basis is
\begin{equation*}
 |\bar j_\beta\rangle
 =V|j\rangle
 =\sum_{k=0}^{d-1}
 (G^{-1/2})_{kj}|w_{k,\beta}\rangle,
 \qquad
 \langle\bar j_\beta|\bar k_\beta\rangle
 =\delta_{jk}.
\end{equation*}
A logical state
\begin{equation*}
 |\psi\rangle=\sum_j\psi_j|j\rangle
\end{equation*}
defines the physical single-mode probe
\begin{equation*}
 |\psi_{\rm GKP}\rangle
 =V|\psi\rangle
 =\sum_j\psi_j|\bar j_\beta\rangle.
\end{equation*}

The maximally entangled two-mode GKP state is
\begin{equation}
 |\Phi_{d,\beta}\rangle
 =\frac1{\sqrt d}
 \sum_{j=0}^{d-1}
 |\bar j_\beta\rangle_A
 |\bar j_\beta\rangle_R.
 \label{eq:finite-bell-state-prelim}
\end{equation}
Its two-mode wavefunction is
\begin{equation*}
 \Psi_{\Phi,\beta}(q_A,q_R)
 =\frac1{\sqrt d}
 \sum_{j=0}^{d-1}
 \bar\psi_{j,\beta}(q_A)
 \bar\psi_{j,\beta}(q_R),
\end{equation*}
where
\begin{equation*}
 \bar\psi_{j,\beta}(q)
 =\langle q|\bar j_\beta\rangle.
\end{equation*}
In the ideal limit,
\begin{equation*}
 |\Phi_{d,L}\rangle
 \propto
 \sum_{j=0}^{d-1}
 \sum_{m,n\in\mathbb Z}
 |q_A=(j+dm)\ell_d\rangle
 |q_R=(j+dn)\ell_d\rangle.
\end{equation*}
The two combs therefore share the same logical residue $j$ modulo $d$. In addition to the local stabilizers, the state is stabilized by
\begin{equation*}
 \bar X_A\bar X_R,
 \qquad
 \bar Z_A\bar Z_R^{-1},
\end{equation*}
which are the modular GKP analogues of the commuting EPR observables
$\hat p_A+\hat p_R$ and $\hat q_A-\hat q_R$.

For a general pure two-qudit logical state $|\Psi\rangle_{AR}$, let
\begin{equation*}
 \tau_A=\Tr_R\ketbra{\Psi}{\Psi}
\end{equation*}
be the signal marginal. When only the signal mode is displaced,
\begin{equation*}
 \kappa_\Psi(\bm\xi)
 =\Tr[
 \tau_AK_\beta^{(d)}(\bm\xi)
 ].
\end{equation*}
For the maximally entangled state in Eq.~\eqref{eq:finite-bell-state-prelim},
\begin{equation*}
 \tau_A=\frac{\id_d}{d},
 \qquad
 \kappa_{\Phi_d}(\bm\xi)
 =\frac1d\Tr K_\beta^{(d)}(\bm\xi).
\end{equation*}

To resolve the complete logical response, define the orthonormal finite-energy Bell basis
\begin{align}
 |\Phi_{ab}^{(\beta)}\rangle
 &=(VW_{ab}\otimes V)|\Phi_d\rangle
 \notag\\
 &=\frac1{\sqrt d}
 \sum_{j=0}^{d-1}
 e^{2\pi ibj/d}
 |\overline{j+a}_\beta\rangle_A
 |\bar j_\beta\rangle_R.
 \label{eq:finite-bell-basis}
\end{align}
Projecting the displaced state onto the encoded two-mode subspace gives
\begin{align*}
 &(P_AD_A(\bm\xi)\otimes P_R)
 |\Phi_{d,\beta}\rangle
 \\
 &\qquad=
 \sum_{a,b=0}^{d-1}
 c_{ab}(\bm\xi)
 |\Phi_{ab}^{(\beta)}\rangle,
\end{align*}
where
\begin{equation}
 c_{ab}(\bm\xi)
 =\frac1d
 \Tr[
 W_{ab}^\dagger
 K_\beta^{(d)}(\bm\xi)
 ].
 \label{eq:bell-expansion-coefficients}
\end{equation}
Weyl orthogonality gives
\begin{equation}
 \sum_{a,b}
 |c_{ab}(\bm\xi)|^2
 =
 \frac1d
 \Tr[
 K_\beta^{(d)\dagger}(\bm\xi)
 K_\beta^{(d)}(\bm\xi)
 ].
 \label{eq:bell-in-code-weight}
\end{equation}
The corresponding leakage probability is
\begin{equation}
 p_{\rm leak}^{\Phi}(\bm\xi)
 =
 1-
 \frac1d
 \Tr[
 K_\beta^{(d)\dagger}(\bm\xi)
 K_\beta^{(d)}(\bm\xi)
 ].
 \label{eq:bell-leakage}
\end{equation}
The identity-Bell amplitude
\begin{equation*}
 c_{00}(\bm\xi)
 =\frac1d\Tr K_\beta^{(d)}(\bm\xi)
\end{equation*}
controls pure identity-versus-displacement discrimination, whereas the full set $\{c_{ab}\}$ is required when random displacements generate mixtures over logical Bell sectors.
% ================================================================
\section{Bayesian approach to process detection in the GKP framework}
\label{sec:bayes}

We consider the process hypotheses
\begin{equation*}
 H_0:\cE_0=\cN_\eta,
 \qquad
 H_1:\cE_1(t)=\cN_\eta\circ\cU_{t\bm u},
\end{equation*}
where
$\cU_{\bm\xi}(\rho)=D(\bm\xi)\rho D^\dagger(\bm\xi)$,
$t\geq0$ is the perturbation amplitude, and $\norm{\bm u}=1$ specifies its direction. The background channel $\cN_\eta$ is first taken to be the identity and is subsequently chosen as a loss--amplifier channel. The Bayesian prior distinguishes the two process hypotheses, while $t$ is scanned as the physical parameter to be detected.

\subsection{Single-mode finite-energy GKP probes}

Let
$\bm c=(c_0,\ldots,c_{d-1})^T$,
with $\bm c^\dagger\bm c=1$, and prepare
\begin{equation*}
 |\psi_{\bm c}\rangle=V\bm c.
\end{equation*}
In the absence of noise, the two possible outputs are
\begin{equation}
 |\psi_0\rangle=V\bm c,
 \qquad
 |\psi_1(t)\rangle=D(t\bm u)V\bm c,
 \label{eq:single-pure-outputs}
\end{equation}
with exact finite-energy overlap
\begin{equation}
 \kappa_{\bm c}(t,\bm u)
 =\bm c^\dagger K_\beta^{(d)}(t\bm u)\bm c
 =\bm c^\dagger G^{-1/2}
 B_\beta(t\bm u)G^{-1/2}\bm c.
 \label{eq:single-kappa-analytic}
\end{equation}
Together, Eqs.~\eqref{eq:explicit-B-double-sum} and
\eqref{eq:single-kappa-analytic} give the complete perturbation dependence directly in terms of the finite-energy GKP comb.

Substitution into Eq.~\eqref{eq:pure-bayes} gives
\begin{equation}
 \Pe^{(1)}(\bm c;t,\bm u)
 =\frac12\left[
 1-\sqrt{
 1-4z_0z_1
 \left|
 \bm c^\dagger
 K_\beta^{(d)}(t\bm u)
 \bm c
 \right|^2
 }
 \right].
 \label{eq:single-gkp-bayes}
\end{equation}
The preparation dependence is therefore contained in the logical matrix element
$\bm c^\dagger K_\beta^{(d)}\bm c$.

In the ideal-code limit, this structure becomes explicit at displacements in the symplectic dual lattice. For
\begin{equation*}
 \bm\xi_{ab}=\ell_d(a,b)^T,
 \qquad
 a,b\in\mathbb Z_d,
\end{equation*}
the logical action is
\begin{equation*}
 K(\bm\xi_{ab})
 =e^{i\phi_{ab}}X^aZ^b,
\end{equation*}
and hence
\begin{equation}
 \kappa_{\bm c}(\bm\xi_{ab})
 =e^{i\phi_{ab}}
 \sum_{j=0}^{d-1}
 c_{j+a}^*c_j
 e^{2\pi ibj/d},
 \label{eq:single-logical-overlap}
\end{equation}
where the indices are understood modulo $d$. A computational state is insensitive to $Z$-type logical shifts and orthogonal to nontrivial $X$-type shifts, whereas a Fourier state has the opposite response. The logical preparation can therefore generate blind directions in displacement space.

For a small physical displacement, define
\begin{equation*}
 \hat G_{\bm u}=u_q\hat p-u_p\hat q,
 \qquad
 D(t\bm u)=e^{-it\hat G_{\bm u}}.
\end{equation*}
A cumulant expansion gives
\begin{align*}
 |\kappa_{\bm c}(t,\bm u)|^2
 &=1-t^2\mathcal V_{\bm c}(\bm u)+O(t^4),
 \\
 \mathcal V_{\bm c}(\bm u)
 &=\langle\hat G_{\bm u}^2\rangle_{V\bm c}
 -\langle\hat G_{\bm u}\rangle_{V\bm c}^2.
\end{align*}
For equal priors,
\begin{equation}
 \Pe^{(1)}(t)
 =\frac12\left[
 1-|t|\sqrt{\mathcal V_{\bm c}(\bm u)}
 +O(|t|^3)
 \right].
 \label{eq:single-small-displacement-bayes}
\end{equation}
Thus the local response is governed by the physical quadrature variance, while the finite-displacement response retains logical zeros and stabilizer revivals.

\subsection{Two-GKP-mode entangled probes}

Let
$|\Psi\rangle_{AR}=\sum_{j,k}\psi_{jk}|j\rangle_A|k\rangle_R$
be a pure logical two-qudit state, with signal marginal
\begin{equation*}
 \tau_A=\Tr_R|\Psi\rangle\langle\Psi|.
\end{equation*}
Its physical encoding is
\begin{equation*}
 |\Psi_{\rm GKP}\rangle=(V\otimes V)|\Psi\rangle.
\end{equation*}
When only mode $A$ is displaced, the overlap between the two hypotheses is
\begin{equation}
 \kappa_\Psi(t,\bm u)
 =\langle\Psi_{\rm GKP}|
 D_A(t\bm u)\otimes\id_R
 |\Psi_{\rm GKP}\rangle
 =\Tr\!\left[
 \tau_AK_\beta^{(d)}(t\bm u)
 \right].
 \label{eq:general-entangled-overlap}
\end{equation}
Equation~\eqref{eq:pure-bayes} then gives
\begin{equation}
 \Pe^{(2)}(\tau_A;t,\bm u)
 =\frac12\left[
 1-\sqrt{
 1-4z_0z_1
 \left|
 \Tr[
 \tau_AK_\beta^{(d)}(t\bm u)
 ]
 \right|^2
 }
 \right].
 \label{eq:general-entangled-bayes}
\end{equation}

For the maximally entangled finite-energy state
$|\Phi_{d,\beta}\rangle$ defined in
Eq.~\eqref{eq:finite-bell-state-prelim},
the signal marginal is $\tau_A=\id_d/d$, and therefore
\begin{align}
 \kappa_{\Phi_d}(t,\bm u)
 &=\frac1d
 \Tr K_\beta^{(d)}(t\bm u),
 \label{eq:bell-overlap-bayes}\\
 \Pe^{\Phi_d}(t)
 &=\frac12\left[
 1-\sqrt{
 1-4z_0z_1
 \left|
 \frac1d
 \Tr K_\beta^{(d)}(t\bm u)
 \right|^2
 }
 \right].
 \label{eq:bell-pure-bayes}
\end{align}
The finite-energy decomposition into logical Bell sectors and the associated leakage probability are given by
Eqs.~\eqref{eq:bell-expansion-coefficients}--\eqref{eq:bell-leakage}.

At an ideal logical displacement
$K=e^{i\phi}W_{a_0b_0}$,
\begin{equation}
 (W_{a_0b_0}\otimes\id)|\Phi_d\rangle
 =\frac1{\sqrt d}
 \sum_{j=0}^{d-1}
 e^{2\pi ib_0j/d}
 |j+a_0\rangle_A|j\rangle_R.
 \label{eq:ideal-bell-label-explicit}
\end{equation}
Direct summation gives
\begin{equation*}
 \langle\Phi_{ab}|\Phi_{a'b'}\rangle
 =\delta_{aa'}\delta_{bb'},
 \qquad
 \frac1d\Tr(X^aZ^b)
 =\delta_{a0}\delta_{b0}.
\end{equation*}
Thus every nontrivial logical displacement maps the initial Bell state to an orthogonal logical Bell sector. The idler retains the initial logical residue, while the signal shift and phase determine the two indices $(a,b)$.

This preparation-independent response does not imply a pointwise precision advantage over all single-mode states.

\textbf{Proposition 1 (pointwise one-mode matching).}
\emph{For every finite-energy compressed displacement
$K_\beta^{(d)}(\bm\xi)$, there exists a normalized one-mode logical vector
$\bm c_\star(\bm\xi)$ such that}
\begin{equation}
 \bm c_\star^\dagger
 K_\beta^{(d)}(\bm\xi)
 \bm c_\star
 =\frac1d
 \Tr K_\beta^{(d)}(\bm\xi).
 \label{eq:pointwise-one-mode-matching}
\end{equation}

\emph{Proof.}
The numerical range
\begin{equation*}
 \cW(K)=
 \{
 \bm c^\dagger K\bm c:
 \norm{\bm c}=1
 \}
\end{equation*}
is convex and contains the spectrum of $K$. It therefore contains the convex hull of the eigenvalues, including their arithmetic mean $\Tr K/d$.
\hfill$\square$

At a fixed known displacement, the Bell overlap can therefore be reproduced by a suitable pure one-mode state, and another one-mode preparation may yield a still smaller overlap. The role of entanglement is instead to remove the dependence on an arbitrarily selected logical input and to provide a uniform response over the displacement family.

More generally, a signal marginal $\tau_A$ reproduces the trace response for every displacement only if
\begin{equation}
 \Tr\!\left[
 \left(\tau_A-\frac{\id_d}{d}\right)
 K_\beta^{(d)}(\bm\xi)
 \right]
 =0
 \qquad
 \text{for all }\bm\xi.
 \label{eq:uniform-trace-condition}
\end{equation}
Whenever the family
$\{K_\beta^{(d)}(\bm\xi)\}_{\bm\xi}$
spans the logical matrix algebra,
Eq.~\eqref{eq:uniform-trace-condition} implies
$\tau_A=\id_d/d$ and hence, for a pure bipartite probe, maximal entanglement.

\subsection{Bayesian benchmarks with coherent, squeezed, and twin-beam probes}
\label{subsec:gaussian-bayes}

Let $t\bm u$ denote the signal displacement and impose the equal-squeezing convention of Eq.~\eqref{eq:equal-squeezing-rule}. The exact squared overlaps of the selected noiseless Gaussian probes are
\begin{align}
 F_{\rm coh}(t)
 &=e^{-t^2/2},
 \label{eq:coherent-fidelity-pure}\\
 F_{\rm sq}(t)
 &=\exp\!\left[-\frac{t^2}{2v_s}\right],
 \notag\\
 F_{\rm TWB}(t)
 &=\exp\!\left[
 -\frac{v_s+v_s^{-1}}{4}t^2
 \right].
 \label{eq:twin-fidelity-pure}
\end{align}
Their Bayesian errors follow from
\begin{equation}
 P_{{\rm e},j}^{\rm pure}(t)
 =\frac12\left[
 1-\sqrt{1-4z_0z_1F_j(t)}
 \right],
 \qquad
 j\in\{{\rm coh,sq,TWB}\}.
 \label{eq:gaussian-pure-bayes}
\end{equation}
For a known direction, the aligned squeezed vacuum is the strongest local member of this selected Gaussian set, whereas the twin-beam response is isotropic.

For equal priors, the corresponding GKP Bell error is
\begin{equation}
 P_{{\rm e},{\rm GKP}}^{\Phi_d}(t)
 =\frac12\left[
 1-\sqrt{
 1-
 \left|
 \frac1d
 \Tr K_\beta^{(d)}(t\bm u)
 \right|^2
 }
 \right].
 \label{eq:gkp-bell-equal-prior-bayes}
\end{equation}
We define the simultaneous noiseless advantage
\begin{equation}
 \Delta_{\rm B}^{\rm pure}(t;d,s)
 =
 \min_{j\in\{{\rm coh,sq,TWB}\}}
 P_{{\rm e},j}^{\rm pure}(t)
 -
 P_{{\rm e},{\rm GKP}}^{\Phi_d}(t).
 \label{eq:pure-common-advantage}
\end{equation}
A positive value means that the finite-energy GKP Bell probe has a lower Bayesian error than all three selected Gaussian probes.

\begin{figure*}[t]
\centering
\includegraphics[width=0.99\textwidth]{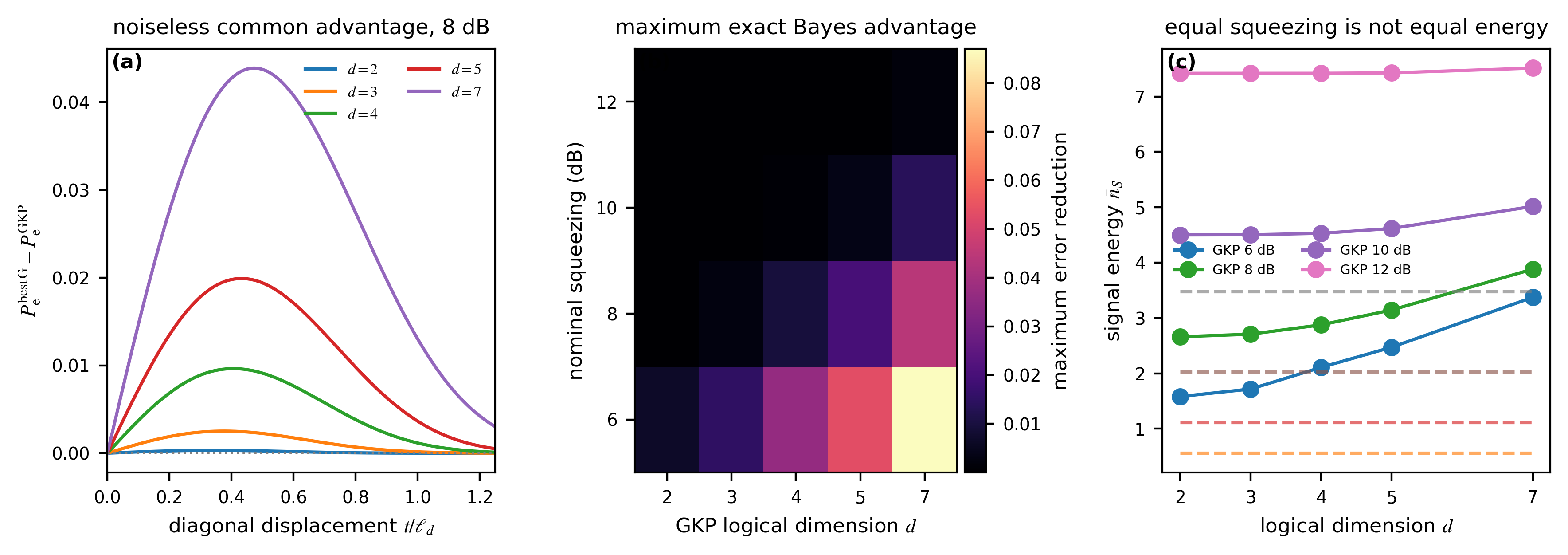}
\caption{Exact noiseless equal-squeezing comparison. (a) Common Bayesian advantage $\Delta_{\rm B}^{\rm pure}$ at $8$ dB for $d=2,3,4,5,7$ along the diagonal phase-space direction. Positive values mean that the finite-energy GKP Bell probe outperforms the coherent, aligned squeezed-vacuum, and twin-beam probes simultaneously. (b) Maximum advantage over $0\leq t/\ell_d\leq1.25$ as a function of logical dimension and nominal squeezing. (c) Code-averaged GKP signal energy (solid curves) and Gaussian signal energy $\sinh^2r$ at the same nominal squeezing (dashed horizontal curves). The comparison is performed at equal squeezing, not equal energy.}
\label{fig:noiseless-equal-squeezing}
\end{figure*}

Figure~\ref{fig:noiseless-equal-squeezing} summarizes the noiseless equal-squeezing comparison. Panel~(a) shows that the GKP advantage is nonlocal: it is negligible near the origin, where the direction-matched squeezed vacuum is locally strong, and becomes positive near the first finite-energy logical feature. Panel~(b) shows how the maximum advantage in the investigated interval varies with logical dimension and nominal squeezing. At $8$ dB, the maximum Bayesian error reduction increases from
$3.29\times10^{-4}$ for $d=2$ to
$4.39\times10^{-2}$ for $d=7$. Panel~(c) reports the corresponding probe energies and emphasizes that equal nominal squeezing does not imply equal photon number.

After gain-compensated loss, homodyne measurement of the displaced quadrature produces equal-variance Gaussian outcomes under the two hypotheses. The coherent, squeezed, and conditional twin-beam variances are
\begin{equation}
 \nu_{\rm coh}
 =\sigma^2+\frac12,
 \qquad
 \nu_{\rm sq}
 =\sigma^2+\frac{v_s}{2},
 \qquad
 \nu_{\rm TWB}
 =\sigma^2+\frac{v_s}{1+v_s^2}.
 \label{eq:noisy-gaussian-receiver-variances}
\end{equation}
For the twin-beam probe, the optimal linear statistic is
\begin{equation*}
 \hat X_g
 =\bm u^T\hat{\bm R}_A
 -g\bm u^TZ\hat{\bm R}_R,
 \qquad
 g_\star=\tanh2r,
\end{equation*}
which gives the last variance in
Eq.~\eqref{eq:noisy-gaussian-receiver-variances}. The resulting classical tests distinguish
$\mathcal N(0,\nu_j)$ from $\mathcal N(t,\nu_j)$ and, for equal priors, achieve
\begin{equation}
 P_{{\rm e},j}^{\rm hom}(t)
 =\Phi_{\rm N}\!\left(
 -\frac{t}{2\sqrt{\nu_j}}
 \right).
 \label{eq:gaussian-homodyne-error}
\end{equation}

We also consider inverse-preparation vacuum-or-not receivers. If the state obtained after undoing the Gaussian preparation has covariance
$\widetilde V_j$ and mean
$\widetilde{\bm d}_{j,h}$ under hypothesis $H_h$, its no-click probability is
\begin{equation}
 q_{j,h}
 =
 \frac{
 \exp\!\left[
 -\widetilde{\bm d}_{j,h}^{T}
 (\widetilde V_j+\id/2)^{-1}
 \widetilde{\bm d}_{j,h}/2
 \right]
 }{
 \sqrt{\det(\widetilde V_j+\id/2)}
 }.
 \label{eq:nulling-off-probability}
\end{equation}
The equal-prior error is
\begin{equation}
 P_{{\rm e},j}^{\rm off}
 =\frac12
 \left[
 1-|q_{j,0}-q_{j,1}|
 \right].
 \label{eq:gaussian-off-error}
\end{equation}
The explicit coherent, squeezed, and twin-beam matrices are given in Appendix~\ref{app:gaussian-nulling}. In the noisy comparisons, the Gaussian benchmark is the pointwise minimum over the three quadrature receivers and the three inverse-preparation vacuum-or-not receivers.

\subsection{Bayesian strategy in the presence of loss and amplification}

We now apply the channel relations in
Eqs.~\eqref{eq:loss-chi}--\eqref{eq:additive-channel}.
If the displacement precedes an uncompensated lossy channel, the output characteristic functions are
\begin{equation}
 \chi_h^{\rm loss}(\bm k)
 =
 e^{-i\sqrt\eta\,\bm k^T\Omega\bm\xi_h}
 \chi_\rho(\sqrt\eta\,\bm k)
 e^{-(1-\eta)\norm{\bm k}^2/4},
 \label{eq:direct-loss-hypotheses}
\end{equation}
where
$\bm\xi_0=0$ and $\bm\xi_1=t\bm u$.
Thus direct loss contracts the phase-space structure of the probe and attenuates the displacement by $\sqrt\eta$.

Post-amplification with gain $G=1/\eta$ gives the unit-gain channel
\begin{equation}
 \cA_{1/\eta}\circ\cL_\eta
 =\cG_{\sigma_{\rm post}^2},
 \qquad
 \sigma_{\rm post}^2
 =\frac{1-\eta}{\eta},
 \label{eq:post-amplified-noise}
\end{equation}
whereas amplification before loss gives
\begin{equation}
 \cL_\eta\circ\cA_{1/\eta}
 =\cG_{\sigma_{\rm pre}^2},
 \qquad
 \sigma_{\rm pre}^2=1-\eta.
 \label{eq:pre-amplified-noise}
\end{equation}
For either unit-gain compensated channel, the two hypotheses correspond to the Gaussian shift densities
\begin{equation}
 g_h(\bm\nu)
 =\frac{1}{2\pi\sigma^2}
 \exp\!\left[
 -\frac{\norm{\bm\nu-\bm\xi_h}^2}{2\sigma^2}
 \right].
 \label{eq:gaussian-shift-laws}
\end{equation}

For an ideal square GKP code, every physical shift can be decomposed as
\begin{equation}
 \bm\nu
 =
 \bm s+\ell_d
 \begin{pmatrix}
 a+dm\\
 b+dn
 \end{pmatrix},
 \qquad
 \bm s\in\cV_d,
 \label{eq:gkp-shift-decomposition}
\end{equation}
where
$(a,b)\in\mathbb Z_d^2$ is the logical Weyl label and
$(m,n)\in\mathbb Z^2$ is a stabilizer translation. The joint density of the analog syndrome and logical label is
\begin{equation}
 f_h^{ab}(\bm s;t)
 =
 \sum_{m,n\in\mathbb Z}
 g_h\!\left[
 \bm s+\ell_d
 \begin{pmatrix}
 a+dm\\
 b+dn
 \end{pmatrix}
 \right].
 \label{eq:wrapped-density-bayes}
\end{equation}
For isotropic noise, this density factorizes as
$f_h^{ab}=F_{h,q}^aF_{h,p}^b$. With
$L_d=d\ell_d=\sqrt{2\pi d}$,
\begin{equation}
 F_{h,\mu}^a(s)
 =\frac1{L_d}
 \vartheta_3\!\left(
 \frac{\pi(s+a\ell_d-\xi_{h,\mu})}{L_d},
 e^{-2\pi^2\sigma^2/L_d^2}
 \right).
 \label{eq:wrapped-theta-factor}
\end{equation}

For a single-mode logical input $\rho_\psi$, ideal syndrome extraction produces the classical--quantum state
\begin{equation}
 \Gamma_h^\psi
 =
 \int_{\cV_d}\dd^2\bm s\,
 \ketbra{\bm s}{\bm s}
 \otimes
 \sum_{a,b}
 f_h^{ab}(\bm s)
 W_{ab}\rho_\psi W_{ab}^\dagger.
 \label{eq:single-cq-output}
\end{equation}
For the GKP Bell probe,
\begin{equation*}
 |\Phi_{ab}\rangle
 =(W_{ab}\otimes\id)|\Phi_d\rangle
\end{equation*}
are mutually orthogonal, and
\begin{equation}
 \Gamma_h^\Phi
 =
 \int_{\cV_d}\dd^2\bm s
 \sum_{a,b}
 f_h^{ab}(\bm s)
 \ketbra{\bm s}{\bm s}
 \otimes
 \ketbra{\Phi_{ab}}{\Phi_{ab}}.
 \label{eq:bell-cq-output}
\end{equation}

To avoid confusion with the Bell amplitudes
$c_{ab}(\bm\xi)$ defined in
Eq.~\eqref{eq:bell-expansion-coefficients}, define the signed likelihoods
\begin{equation}
 \ell_{ab}(\bm s;t)
 =
 z_1f_1^{ab}(\bm s;t)
 -z_0f_0^{ab}(\bm s).
 \label{eq:signed-wrapped-likelihood}
\end{equation}
The Bell Bayesian operator is diagonal in the joint syndrome--Bell basis, so
\begin{equation}
 \Pe^\Phi(t)
 =\frac12\left[
 1-
 \sum_{a,b}
 \int_{\cV_d}\dd^2\bm s\,
 |\ell_{ab}(\bm s;t)|
 \right].
 \label{eq:bell-bayes-noisy}
\end{equation}

For an arbitrary single-mode input,
\begin{equation}
 \Delta_\psi(\bm s;t)
 =
 \sum_{a,b}
 \ell_{ab}(\bm s;t)
 W_{ab}\rho_\psi W_{ab}^\dagger,
 \label{eq:single-syndrome-bayesian-block}
\end{equation}
and the optimized error is
\begin{equation}
 \Pe^{(1),\star}(t)
 =
 \min_{\psi}
 \frac12\left[
 1-
 \int_{\cV_d}\dd^2\bm s\,
 \norm{\Delta_\psi(\bm s;t)}_1
 \right].
 \label{eq:product-bayes-general}
\end{equation}

Two important single-mode preparations are diagonal for arbitrary $d$. For the computational GKP state $|0\rangle$, the phase label $b$ is irrelevant. Define
\begin{equation}
 q_{h,a}^{(Z)}(\bm s;t)
 =
 \sum_{b=0}^{d-1}
 f_h^{ab}(\bm s;t).
 \label{eq:computational-marginal-density}
\end{equation}
The Bayesian block is
\begin{align*}
 \Delta_Z(\bm s;t)
 &=\sum_{a=0}^{d-1}
 d_a^{(Z)}(\bm s;t)
 \ketbra{a}{a},
 \\
 d_a^{(Z)}(\bm s;t)
 &=z_1q_{1,a}^{(Z)}(\bm s;t)
 -z_0q_{0,a}^{(Z)}(\bm s),
\end{align*}
and hence
\begin{equation}
 \Pe^{Z}(t)
 =\frac12\left[
 1-
 \int_{\cV_d}\dd^2\bm s
 \sum_{a=0}^{d-1}
 |d_a^{(Z)}(\bm s;t)|
 \right].
 \label{eq:Z-bayes-arbitrary-d}
\end{equation}
For the Fourier GKP state
$|+\rangle=d^{-1/2}\sum_j|j\rangle$,
one instead defines
\begin{equation*}
 q_{h,b}^{(X)}(\bm s;t)
 =\sum_a f_h^{ab}(\bm s;t),
\end{equation*}
and obtains the corresponding expression after
$Z\leftrightarrow X$ and $a\leftrightarrow b$.

For a GKP qubit, Eq.~\eqref{eq:product-bayes-general} can be diagonalized for an arbitrary single-mode preparation. Write
\begin{equation*}
 \rho_{\bm r}
 =\frac12
 (\id+r_xX+r_yY+r_zZ),
 \qquad
 \norm{\bm r}\leq1,
\end{equation*}
and order the logical labels as
$W_{00}=I$,
$W_{10}=X$,
$W_{01}=Z$, and
$W_{11}\sim Y$.
Define the Walsh transforms
\begin{align*}
 C_0
 &=\ell_{00}+\ell_{10}+\ell_{01}+\ell_{11},
 \\
 C_x
 &=\ell_{00}+\ell_{10}-\ell_{01}-\ell_{11},
 \\
 C_y
 &=\ell_{00}-\ell_{10}-\ell_{01}+\ell_{11},
 \\
 C_z
 &=\ell_{00}-\ell_{10}+\ell_{01}-\ell_{11}.
\end{align*}
Then
\begin{equation*}
 \Delta_{\bm r}(\bm s;t)
 =\frac12\left[
 C_0\id
 +r_xC_xX
 +r_yC_yY
 +r_zC_zZ
 \right],
\end{equation*}
with eigenvalues
\begin{equation*}
 \lambda_\pm(\bm s)
 =\frac12
 \left[
 C_0\pm R_{\bm r}
 \right],
 \qquad
 R_{\bm r}
 =
 \sqrt{
 r_x^2C_x^2+
 r_y^2C_y^2+
 r_z^2C_z^2
 }.
\end{equation*}
Therefore
\begin{equation*}
 \norm{\Delta_{\bm r}(\bm s;t)}_1
 =\max\{|C_0|,R_{\bm r}\},
\end{equation*}
and
\begin{equation}
 \Pe^{(1)}(\bm r;t)
 =\frac12\left[
 1-
 \int_{\cV_2}\dd^2\bm s\,
 \max\{|C_0|,R_{\bm r}\}
 \right].
 \label{eq:qubit-bayes-explicit}
\end{equation}
Because
\begin{equation*}
 R_{\bm r}
 \leq\max_{i\in\{x,y,z\}}|C_i|
 \leq\sum_{a,b}|\ell_{ab}|,
 \qquad
 |C_0|
 \leq\sum_{a,b}|\ell_{ab}|,
\end{equation*}
the ideal-code Bell strategy in
Eq.~\eqref{eq:bell-bayes-noisy}
is never worse than the single-mode GKP-qubit strategy in
Eq.~\eqref{eq:qubit-bayes-explicit}.

For axial displacements, the two strategies can coincide in every logical dimension. For example, if $\xi_{h,p}=0$, the signed likelihood factorizes as
\begin{equation*}
 \ell_{ab}(s_q,s_p)
 =B^b(s_p)d_a(s_q),
 \qquad
 B^b(s_p)\geq0.
\end{equation*}
The Bell trace norm is then
\begin{equation*}
 \sum_{a,b}|\ell_{ab}|
 =
 \left(\sum_bB^b\right)
 \left(\sum_a|d_a|\right).
\end{equation*}
For the computational input,
\begin{equation*}
 d_a^{(Z)}
 =
 \left(\sum_bB^b\right)d_a,
\end{equation*}
and therefore
\begin{equation}
 \sum_a|d_a^{(Z)}|
 =\sum_{a,b}|\ell_{ab}|.
 \label{eq:axial-equality}
\end{equation}
Thus a shift along $q$ gives identical Bayesian risks for the computational single-mode probe and the entanglement-assisted probe for every $d$. A Fourier input gives the corresponding equality for a shift along $p$. For oblique displacements, both noncommuting logical labels generally carry hypothesis information, and the inequality may be strict.

For finite-energy GKP states, the ideal syndrome sectors are not exactly orthogonal. The mixed-state calculation can nevertheless be expressed through the same kernel
$K_\beta^{(d)}$. Approximate the Gaussian displacement law by positive quadrature nodes
$\{\bm\nu_{hr},w_{hr}\}$ and define
\begin{equation*}
 |\Psi_{hr}^{(\tau)}\rangle
 =
 (D(\bm\nu_{hr})V\otimes\id)
 |\Psi_\tau\rangle,
\end{equation*}
where $|\Psi_\tau\rangle$ purifies the signal marginal $\tau$. The pairwise overlaps are
\begin{equation}
 \langle\Psi_{hr}^{(\tau)}
 |\Psi_{ks}^{(\tau)}\rangle
 =
 e^{i\bm\nu_{hr}^T\Omega\bm\nu_{ks}/2}
 \Tr\!\left[
 \tau
 K_\beta^{(d)}
 (\bm\nu_{ks}-\bm\nu_{hr})
 \right].
 \label{eq:finite-orbit-kernel}
\end{equation}
Let $G_z$ be the Gram matrix of the weighted vectors
\begin{equation*}
 \sqrt{z_hw_{hr}}
 |\Psi_{hr}^{(\tau)}\rangle
\end{equation*}
and define
\begin{equation*}
 J=\operatorname{diag}
 (-\id_{N_0},+\id_{N_1}).
\end{equation*}
The nonzero spectrum of the Helstrom operator for the discretized mixture is the spectrum of
\begin{equation*}
 B_z=G_z^{1/2}JG_z^{1/2},
\end{equation*}
so that
\begin{equation}
 \Pe^{(\tau)}
 =\frac12
 \left(
 1-\norm{B_z}_1
 \right).
 \label{eq:finite-energy-gram-bayes}
\end{equation}
Finite squeezing, nonorthogonality, leakage, loss, and amplification enter through the theta kernel in
Eq.~\eqref{eq:finite-orbit-kernel}. The continuous Gaussian mixture is recovered by increasing the quadrature order.

For the comparative parameter sweep, the finite-energy mixed output is constructed directly. Since the idler is noiseless, its physical GKP encoding may be replaced by a $d$-dimensional reference through a local isometry, without changing the trace norms or the spectra of the decision operators. Define
\begin{equation*}
 \Phi_{d,\beta}
 =
 |\Phi_{d,\beta}\rangle
 \langle\Phi_{d,\beta}|.
\end{equation*}
The null and displaced states are
\begin{align}
 \rho_0^{\Phi_d}
 &=
 \int
 \frac{\dd^2\bm\nu}{2\pi\sigma^2}
 e^{-\norm{\bm\nu}^2/(2\sigma^2)}
 \notag\\[-0.5ex]
 &\quad\times
 (D_A(\bm\nu)\otimes\id_R)
 \Phi_{d,\beta}
 (D_A^\dagger(\bm\nu)\otimes\id_R),
 \label{eq:finite-gkp-null-output}\\
 \rho_1^{\Phi_d}(t)
 &=
 (D_A(t\bm u)\otimes\id_R)
 \rho_0^{\Phi_d}
 (D_A^\dagger(t\bm u)\otimes\id_R).
 \label{eq:finite-gkp-alternative-output}
\end{align}
The calculations are verified for convergence with respect to the Fock-space truncation and Gaussian quadrature order.

\begin{figure*}[t]
\centering
\includegraphics[width=0.99\textwidth]{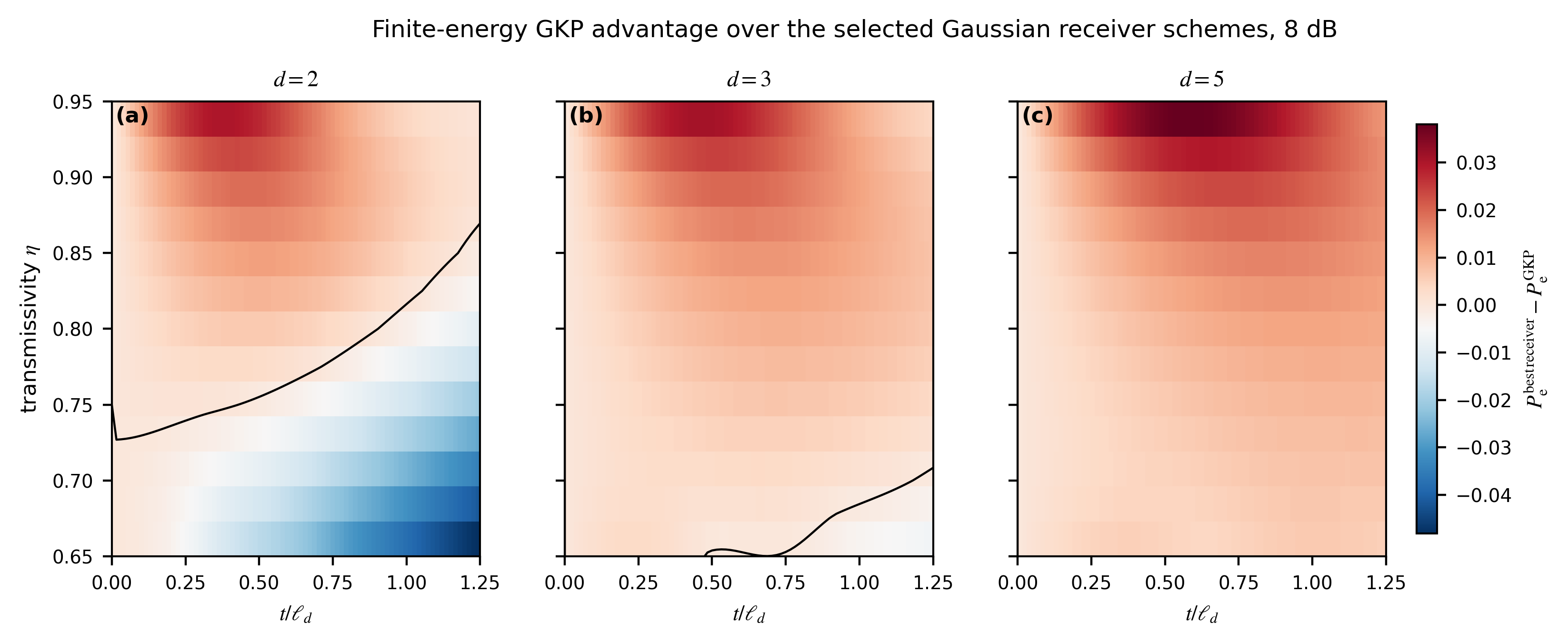}
\caption{Finite-energy Bayesian comparison after loss and post-amplification at $8$ dB. The color represents $\Delta_{\rm B}^{\rm rec}$ from Eq.~\eqref{eq:noisy-common-advantage}; red denotes a lower GKP Helstrom error than every selected Gaussian receiver, blue denotes the opposite, and the black curves mark equality. The Gaussian receiver set comprises coherent- and squeezed-state homodyne detection, twin-beam conditional homodyne detection, and the three inverse-preparation vacuum-or-not receivers.}
\label{fig:noisy-common-advantage}
\end{figure*}

Finally, define the noisy receiver advantage
\begin{equation}
 \Delta_{\rm B}^{\rm rec}(t;d,s,\eta)
 =
 P_{\rm e}^{\rm best\,selected\,receiver}(t)
 -
 P_{{\rm e},{\rm GKP}}^\star(t).
 \label{eq:noisy-common-advantage}
\end{equation}
Figure~\ref{fig:noisy-common-advantage} maps this quantity after loss and post-amplification at $8$ dB. Positive regions identify displacement and transmissivity values for which the finite-energy GKP protocol has a lower error than every selected Gaussian receiver. The largest reduction in the investigated regime is
$0.03809$ at
$d=5$, $\eta=0.95$, and
$t/\ell_5=0.578125$, where
\begin{equation*}
 P_{{\rm e},{\rm GKP}}^\star=0.14802,
 \qquad
 P_{\rm e}^{\rm best\,selected\,receiver}=0.18611.
\end{equation*}
At $\eta=0.8$, the maximum reductions are
$0.00641$, $0.01024$, and $0.01205$ for
$d=2,3,5$, respectively.

%%%%%%%%%%%%%%%%%%%%%%%%%%%%%%%%%%%%%%%%%%%%%%%%%%%%%%%%%%%%
%%%%%%%%%
%%%%%%%%%%%%%%%%

\section{Neyman--Pearson strategy and the minimum detectable perturbation}
\label{sec:np}

\subsection{Pure finite-energy GKP probes}

For the one-mode probe, define
\begin{equation*}
 k_{\bm c}(t)
 =|\kappa_{\bm c}(t,\bm u)|
 =\left|
 \bm c^\dagger K_\beta^{(d)}(t\bm u)\bm c
 \right|.
\end{equation*}
For the two-GKP Bell probe, the corresponding overlap is
\begin{equation*}
 k_\Phi(t)
 =\left|
 \frac1d\Tr K_\beta^{(d)}(t\bm u)
 \right|.
\end{equation*}
Writing $k_\omega(t)$ for either $k_{\bm c}(t)$ or $k_\Phi(t)$, the pure-state result in Eq.~\eqref{eq:pure-roc-prelim} gives
\begin{equation*}
 p_{11}^{\omega}(p_{10};t)=
 \begin{cases}
 \left[
 \sqrt{p_{10}}\,k_\omega(t)
 +\sqrt{1-p_{10}}\sqrt{1-k_\omega^2(t)}
 \right]^2,
 &0\leq p_{10}\leq k_\omega^2(t),\\[1mm]
 1,
 &k_\omega^2(t)<p_{10}\leq1.
 \end{cases}
\end{equation*}
The corresponding optimal receiver is obtained from
Eqs.~\eqref{eq:pure-np-eigs} and
\eqref{eq:pure-np-eigenvector} by substituting the appropriate physical GKP overlap. Proposition~1 implies that, at each fixed displacement, the Bell overlap can be reproduced by a suitable one-mode state, although a fixed one-mode preparation may remain blind along a logical direction.

Define the discrimination power
\begin{equation*}
 \mathcal A_\omega(t)
 =1-|\kappa_\omega(t)|^2.
\end{equation*}
On the nontrivial ROC branch, let
\begin{equation*}
 p_{10}=\alpha=\sin^2\theta,
 \qquad
 \mathcal A_\omega(t)=\sin^2\varphi.
\end{equation*}
Since
$p_{11}=\sin^2(\theta+\varphi)$,
the target condition
$p_{11}\geq\zeta$, with $\alpha\leq\zeta$, is equivalent to
\begin{align*}
 \mathcal A_\omega(t)
 &\geq a_\zeta(\alpha),\\
 a_\zeta(\alpha)
 &=
 \left[
 \sqrt{\zeta(1-\alpha)}
 -\sqrt{(1-\zeta)\alpha}
 \right]^2.
\end{align*}
For the absolute criterion $\zeta=1/2$,
\begin{equation*}
 a_{\rm abs}(\alpha)
 =\frac12-\sqrt{\alpha(1-\alpha)},
 \qquad
 0\leq\alpha\leq\frac12.
\end{equation*}
The relative criterion
$p_{11}/p_{10}\geq\delta$
is obtained by setting $\zeta=\delta\alpha$:
\begin{equation*}
 a_{\rm rel}(\alpha,\delta)
 =
 \alpha
 \left[
 \sqrt{\delta(1-\alpha)}
 -\sqrt{1-\delta\alpha}
 \right]^2,
 \qquad
 \delta\alpha\leq1.
\end{equation*}

For the finite-energy GKP probes,
\begin{align*}
 \mathcal A_{\bm c}(t)
 &=1-
 \left|
 \bm c^\dagger
 G^{-1/2}B_\beta(t\bm u)G^{-1/2}
 \bm c
 \right|^2,\\
 \mathcal A_\Phi(t)
 &=1-
 \frac1{d^2}
 \left|
 \Tr\!\left[
 G^{-1/2}B_\beta(t\bm u)G^{-1/2}
 \right]
 \right|^2.
\end{align*}
The first detectable displacement for a fixed probe $\omega$ is therefore
\begin{equation*}
 t_{\rm m}^{\omega}
 (\bm u,\alpha,\zeta)
 =
 \inf
 \left\{
 t\geq0:
 \mathcal A_\omega(t)
 \geq a_\zeta(\alpha)
 \right\}.
\end{equation*}
The first-crossing convention is required because logical minima may be followed by stabilizer revivals.

A cumulant expansion gives a controlled local approximation. Let
$\kappa_n(\hat G_{\bm u})$ denote the cumulants of
\begin{equation*}
 \hat G_{\bm u}
 =u_q\hat p-u_p\hat q
\end{equation*}
in the selected probe, and write
$V_\omega=\kappa_2$. Since
\begin{equation*}
 \log
 \left|
 \left\langle
 e^{-it\hat G_{\bm u}}
 \right\rangle
 \right|^2
 =
 -V_\omega t^2
 +\frac{\kappa_4}{12}t^4
 +O(t^6),
\end{equation*}
one obtains
\begin{equation*}
 \mathcal A_\omega(t)
 =
 V_\omega t^2
 -
 \left(
 \frac{V_\omega^2}{2}
 +\frac{\kappa_4}{12}
 \right)t^4
 +O(t^6).
\end{equation*}
Perturbative inversion yields
\begin{equation}
 t_{\rm m}^{\rm loc}
 =
 \sqrt{\frac{a_\zeta}{V_\omega}}
 \left[
 1+
 \frac{a_\zeta}{2V_\omega^2}
 \left(
 \frac{V_\omega^2}{2}
 +\frac{\kappa_4}{12}
 \right)
 +O(a_\zeta^2)
 \right].
 \label{eq:tmin-local}
\end{equation}
For the two-mode GKP Bell probe,
\begin{equation*}
 V_\Phi(\bm u)
 =
 \frac1d\Tr[P\hat G_{\bm u}^2]
 -
 \left(
 \frac1d\Tr[P\hat G_{\bm u}]
 \right)^2,
 \qquad
 P=VV^\dagger,
\end{equation*}
with higher cumulants obtained from the corresponding code-averaged moments.

\subsection{Neyman--Pearson benchmarks for the three Gaussian probes}
\label{subsec:gaussian-np}

For the noiseless coherent, squeezed-vacuum, and TMSV probes, substitution of
Eqs.~\eqref{eq:coherent-fidelity-pure}--\eqref{eq:twin-fidelity-pure}
into the pure-state ROC gives
\begin{equation*}
 P_{{\rm D},j}^{\star}(\alpha;t)
 =
 \begin{cases}
 \left[
 \sqrt{\alpha F_j(t)}
 +\sqrt{(1-\alpha)[1-F_j(t)]}
 \right]^2,
 &\alpha\leq F_j(t),\\[1mm]
 1,
 &\alpha>F_j(t),
 \end{cases}
\end{equation*}
where
$j\in\{{\rm coh,sq,TWB}\}$.
Writing
\begin{equation*}
 F_j(t)
 =\exp\!\left[
 -\frac{t^2}{4\nu_j^{(0)}}
 \right],
\end{equation*}
the exact noiseless first-crossing threshold is
\begin{equation*}
 t_{{\rm m},j}^{\rm pure}
 =
 2\sqrt{\nu_j^{(0)}}
 \left[
 \ln\frac{1}{1-a_\zeta(\alpha)}
 \right]^{1/2},
\end{equation*}
with
\begin{equation*}
 \nu_{\rm coh}^{(0)}=\frac12,
 \qquad
 \nu_{\rm sq}^{(0)}=\frac{v_s}{2},
 \qquad
 \nu_{\rm TWB}^{(0)}
 =\frac{1}{v_s+v_s^{-1}}.
\end{equation*}

After loss and amplification, the quadrature-receiver outcomes have means $0$ and $t$ and variances $\nu_j$ from
Eq.~\eqref{eq:noisy-gaussian-receiver-variances}. Their likelihood ratio is monotone, giving
\begin{align}
 P_{{\rm D},j}^{\rm hom}(\alpha;t)
 &=
 \Phi_{\rm N}\!\left[
 \Phi_{\rm N}^{-1}(\alpha)
 +\frac{t}{\sqrt{\nu_j}}
 \right],
 \label{eq:gaussian-receiver-roc}\\
 t_{{\rm m},j}^{\rm hom}(\alpha,\zeta)
 &=
 \sqrt{\nu_j}
 \left[
 \Phi_{\rm N}^{-1}(\zeta)
 -\Phi_{\rm N}^{-1}(\alpha)
 \right].
 \notag
\end{align}

For an inverse-preparation vacuum-or-not receiver, let
\begin{equation*}
 c_h=1-q_h
\end{equation*}
be the click probability under $H_h$. Since the displacement lowers the vacuum probability, the click outcome is tested first. Randomization on the two outcomes gives the complete ROC
\begin{equation}
 P_{{\rm D},j}^{\rm off}(\alpha;t)
 =
 \begin{cases}
 \alpha\,c_1/c_0,
 &0\leq\alpha\leq c_0,\\[1mm]
 c_1+(\alpha-c_0)q_1/q_0,
 &c_0<\alpha\leq1.
 \end{cases}
 \label{eq:nulling-np-roc}
\end{equation}
Equations~\eqref{eq:gaussian-receiver-roc} and
\eqref{eq:nulling-np-roc} define the six selected receiver benchmarks. The exact finite-energy GKP ROC is evaluated using the Gram-matrix dual derived below.

\begin{figure*}[t]
\centering
\includegraphics[width=0.99\textwidth]{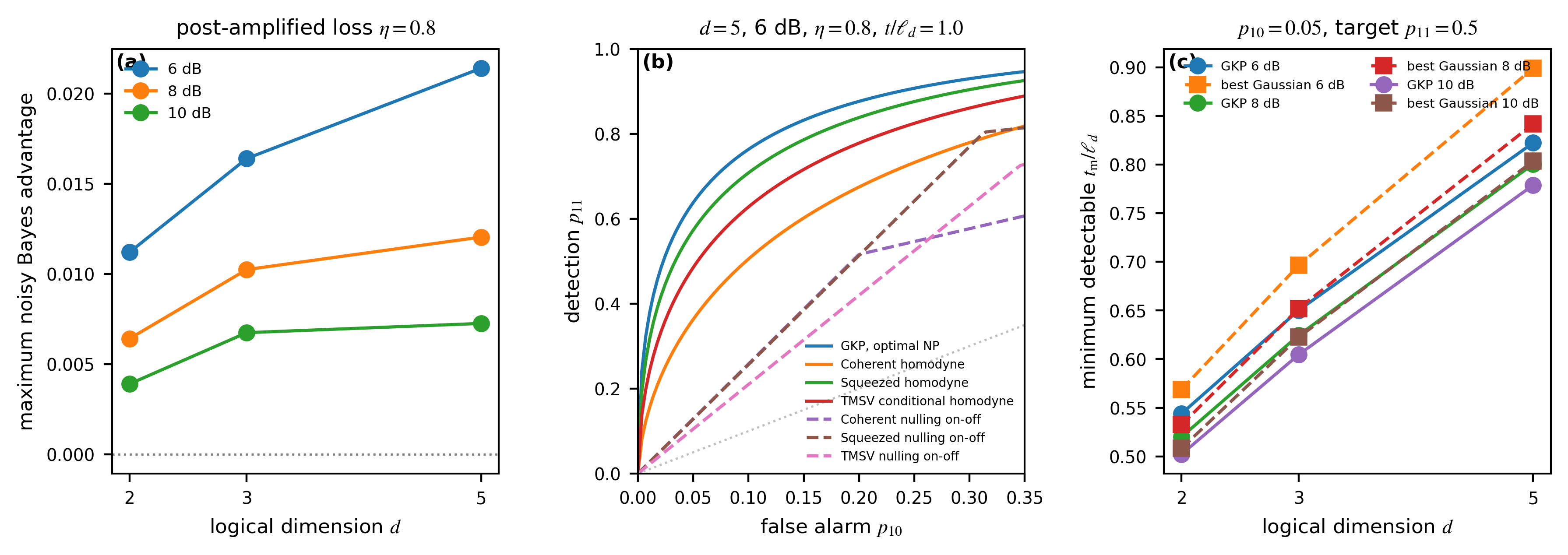}
\caption{Selected-receiver Bayesian and Neyman--Pearson comparison. (a) Maximum noisy Bayesian advantage at $\eta=0.8$ versus logical dimension for $6$, $8$, and $10$ dB. (b) Complete ROC at $d=5$, $6$ dB, $\eta=0.8$, and $t/\ell_5=1$. Solid curves show the optimal GKP test and the Gaussian quadrature receivers; dashed curves show the inverse-preparation vacuum-or-not receivers. (c) First-crossing minimum detectable displacement at $p_{10}=0.05$ and target $p_{11}=1/2$. The Gaussian benchmark is the best of the six displayed receiver schemes at each squeezing.}
\label{fig:receiver-np-comparison}
\end{figure*}

Figure~\ref{fig:receiver-np-comparison} summarizes the Bayesian and Neyman--Pearson comparisons. At
$d=5$, $6$ dB, $\eta=0.8$, and
$t/\ell_5=1$, the exact GKP detection probability at
$p_{10}=0.05$ is $0.63801$, compared with
$0.57310$ for squeezed homodyne,
$0.48512$ for TMSV conditional homodyne, and
$0.36301$ for coherent homodyne. The inverse-preparation on--off receivers are weaker at this operating point. For the target
$p_{11}=1/2$, the thresholds are
\begin{equation*}
 t_{\rm m}^{\rm GKP}=0.92197,
 \qquad
 t_{\rm m}^{\rm best\,selected}=1.00806,
\end{equation*}
corresponding to an $8.54\%$ reduction. At $d=5$, the reduction remains
$4.91\%$ at $8$ dB and
$3.07\%$ at $10$ dB.

\subsection{Ideal GKP Neyman--Pearson operator under loss and amplification}

For the entanglement-assisted ideal GKP protocol, the Lagrange operator is diagonal in the syndrome--Bell basis:
\begin{equation*}
 \Gamma_\gamma^\Phi(t)
 =
 \int_{\cV_d}\dd^2\bm s
 \sum_{a,b}
 \left[
 f_1^{ab}(\bm s;t)
 -\gamma f_0^{ab}(\bm s)
 \right]
 \ketbra{\bm s,\Phi_{ab}}{\bm s,\Phi_{ab}}.
\end{equation*}
Define the likelihood ratio
\begin{equation*}
 \mathcal R_{ab}(\bm s;t)
 =
 \frac{f_1^{ab}(\bm s;t)}
 {f_0^{ab}(\bm s)}.
\end{equation*}
The optimal test is the classical likelihood-ratio rule on the physical GKP record
$X=(\bm s,a,b)$. Including randomization on a level set of nonzero probability,
\begin{align*}
 Q_{\gamma,\nu}^{\Phi}
 &=
 \int_{\cV_d}\dd^2\bm s
 \sum_{a,b}
 \Bigl[
 \mathbf 1_{\{\mathcal R_{ab}>\gamma\}}
 +\nu\mathbf 1_{\{\mathcal R_{ab}=\gamma\}}
 \Bigr]
 \ketbra{\bm s,\Phi_{ab}}{\bm s,\Phi_{ab}},
 \\
 &\hspace{45mm}0\leq\nu\leq1.
\end{align*}
The corresponding false-alarm and detection probabilities are
\begin{align}
 p_{10}^{\Phi}(\gamma,\nu;t)
 &=
 P_0(\mathcal R>\gamma)
 +\nu P_0(\mathcal R=\gamma),
 \label{eq:bell-p10-param}
 \\
 p_{11}^{\Phi}(\gamma,\nu;t)
 &=
 P_1(\mathcal R>\gamma)
 +\nu P_1(\mathcal R=\gamma).
 \label{eq:bell-p11-param}
\end{align}
Equivalently,
\begin{align*}
 \PD^\Phi(\alpha;t)
 =
 \min_{\gamma\geq0}
 \Bigg\{
 \gamma\alpha
 +\sum_{a,b}
 \int_{\cV_d}\dd^2\bm s\,
 \pos{
 f_1^{ab}(\bm s;t)
 -\gamma f_0^{ab}(\bm s)
 }
 \Bigg\}.
\end{align*}
For the continuous Gaussian model, the equality set has zero measure generically and $\nu$ is unnecessary. At differentiable points, the ROC is concave and
\begin{equation*}
 \frac{\dd\PD^\Phi}{\dd\alpha}
 =\gamma_\alpha.
\end{equation*}

The likelihood ratio has an explicit theta-function representation. Let
\begin{equation*}
 L_d=d\ell_d,
 \qquad
 Q_\sigma
 =\exp[-2\pi^2\sigma^2/L_d^2].
\end{equation*}
Poisson summation gives
\begin{equation*}
 f_{h,q}^{a}(s_q)
 =
 \frac1{L_d}
 \vartheta_3\!\left(
 \frac{\pi[s_q+a\ell_d-\xi_{h,q}]}{L_d},
 Q_\sigma
 \right),
\end{equation*}
with the analogous expression for the $p$ component. Hence
\begin{equation}
 \mathcal R_{ab}(\bm s;t)
 =
 \prod_{\mu=q,p}
 \frac{
 \vartheta_3\!\left(
 \pi[s_\mu+r_\mu\ell_d-tu_\mu]/L_d,
 Q_\sigma
 \right)
 }{
 \vartheta_3\!\left(
 \pi[s_\mu+r_\mu\ell_d]/L_d,
 Q_\sigma
 \right)
 },
 \qquad
 (r_q,r_p)=(a,b).
 \label{eq:theta-likelihood-ratio}
\end{equation}
Equation~\eqref{eq:theta-likelihood-ratio} defines the explicit GKP likelihood-ratio receiver under loss and gain compensation.

If wrapping is ignored and the full displacement sample $\bm\nu$ is observed, the Gaussian log-likelihood ratio is
\begin{equation*}
 \log
 \frac{g_1(\bm\nu;t)}
 {g_0(\bm\nu)}
 =
 \frac{
 t\bm u^T\bm\nu-t^2/2
 }{\sigma^2}.
\end{equation*}
The sufficient statistic is
$Y=\bm u^T\bm\nu$, and the exact unwrapped benchmark is
\begin{align}
 P_D^{\rm unwrapped}(\alpha;t)
 &=
 \Phi_{\rm N}\!\left[
 \Phi_{\rm N}^{-1}(\alpha)
 +\frac{t}{\sigma}
 \right],
 \notag\\
 t_{\rm m}^{\rm unwrapped}(\alpha,\zeta)
 &=
 \sigma
 \left[
 \Phi_{\rm N}^{-1}(\zeta)
 -\Phi_{\rm N}^{-1}(\alpha)
 \right].
 \label{eq:unwrapped-gaussian-tmin}
\end{align}
The wrapped receiver approaches this expression when the noise and mean shift remain well inside one unambiguous cell; deviations quantify logical aliasing.

For small $t$,
\begin{equation*}
 f_t^{ab}(\bm s)
 =
 f_0^{ab}(\bm s)
 \left[
 1+t\mathcal S_{ab}(\bm s)
 +O(t^2)
 \right],
\end{equation*}
where the score is
\begin{equation*}
 \mathcal S_{ab}(\bm s)
 =
 -\bm u\cdot
 \bm\nabla_{\bm s}
 \log f_0^{ab}(\bm s).
\end{equation*}
Its derivatives follow directly from $\vartheta_3'$. Let
$c_\alpha$ satisfy
\begin{equation*}
 P_0(\mathcal S>c_\alpha)=\alpha.
\end{equation*}
The locally most-powerful detection probability is
\begin{equation*}
 P_D^\Phi(\alpha;t)
 =
 \alpha
 +t\,
 \mathbb E_0
 \left[
 \mathcal S
 \mathbf 1_{\{\mathcal S>c_\alpha\}}
 \right]
 +O(t^2),
\end{equation*}
and therefore
\begin{equation}
 t_{\rm m}^{\rm LMP}(\alpha,\zeta)
 =
 \frac{\zeta-\alpha}{
 \mathbb E_0[
 \mathcal S
 \mathbf 1_{\{\mathcal S>c_\alpha\}}
 ]}
 +O[(\zeta-\alpha)^2].
 \label{eq:lmp-tmin}
\end{equation}

For a computational GKP state, the arbitrary-$d$ Lagrange operator is diagonal:
\begin{equation*}
 \Gamma_{\gamma,Z}(\bm s;t)
 =
 \sum_{a=0}^{d-1}
 \left[
 q_{1,a}^{(Z)}(\bm s;t)
 -\gamma q_{0,a}^{(Z)}(\bm s)
 \right]
 \ketbra{a}{a}.
\end{equation*}
Its parametric ROC is
\begin{align*}
 p_{10}^{Z}(\gamma;t)
 &=
 \sum_a
 \int_{
 q_{1,a}^{(Z)}
 >
 \gamma q_{0,a}^{(Z)}
 }
 \dd^2\bm s\,
 q_{0,a}^{(Z)}(\bm s),
 \\
 p_{11}^{Z}(\gamma;t)
 &=
 \sum_a
 \int_{
 q_{1,a}^{(Z)}
 >
 \gamma q_{0,a}^{(Z)}
 }
 \dd^2\bm s\,
 q_{1,a}^{(Z)}(\bm s;t).
\end{align*}
A Fourier input obeys the analogous formulas with
$q_{h,b}^{(X)}=\sum_a f_h^{ab}$. In an axial problem, these single-mode likelihood-ratio tests retain all hypothesis-dependent information and coincide with the Bell test for every $d$.

For a one-mode GKP qubit, the characteristic operator can be diagonalized for an arbitrary logical preparation. Define
\begin{align*}
 F_{h,0}
 &=f_h^{00}+f_h^{10}+f_h^{01}+f_h^{11},\\
 F_{h,x}
 &=f_h^{00}+f_h^{10}-f_h^{01}-f_h^{11},\\
 F_{h,y}
 &=f_h^{00}-f_h^{10}-f_h^{01}+f_h^{11},\\
 F_{h,z}
 &=f_h^{00}-f_h^{10}+f_h^{01}-f_h^{11},
\end{align*}
and
\begin{equation*}
 D_\mu(\bm s;\gamma,t)
 =
 F_{1,\mu}(\bm s;t)
 -\gamma F_{0,\mu}(\bm s),
 \qquad
 \mu=0,x,y,z.
\end{equation*}
For
\begin{equation*}
 \rho_{\bm r}
 =
 \frac12
 \left(
 \id+r_xX+r_yY+r_zZ
 \right),
\end{equation*}
the syndrome-resolved Lagrange operator is
\begin{equation*}
 \Gamma_{\gamma,\bm r}(\bm s;t)
 =
 \frac12
 \left[
 D_0\id
 +r_xD_xX
 +r_yD_yY
 +r_zD_zZ
 \right].
\end{equation*}
Its eigenvalues are
\begin{equation}
 g_\pm(\bm s)
 =
 \frac12
 \left[
 D_0\pm R_\gamma
 \right],
 \qquad
 R_\gamma
 =
 \sqrt{
 r_x^2D_x^2+
 r_y^2D_y^2+
 r_z^2D_z^2
 }.
 \label{eq:qubit-np-block-eigs}
\end{equation}
With
\begin{equation*}
 \bm v_\gamma
 =
 (r_xD_x,r_yD_y,r_zD_z),
 \qquad
 R_\gamma=\norm{\bm v_\gamma},
\end{equation*}
the positive projector is
\begin{equation*}
 Q_{\gamma,\bm r}(\bm s)
 =
 \begin{cases}
 \id,
 &D_0\geq R_\gamma,\\
 0,
 &D_0\leq-R_\gamma,\\
 \tfrac12
 \left[
 \id+
 \bm v_\gamma\cdot\bm\sigma/R_\gamma
 \right],
 &|D_0|<R_\gamma.
 \end{cases}
\end{equation*}
Define
\begin{equation*}
 S_h(\bm s;\gamma,\bm r)
 =
 \sum_{i=x,y,z}
 r_i^2D_iF_{h,i}
\end{equation*}
and
\begin{equation*}
 \mathcal T_h(\bm s;\gamma,\bm r)
 =
 \begin{cases}
 F_{h,0},
 &D_0\geq R_\gamma,\\
 0,
 &D_0\leq-R_\gamma,\\
 \tfrac12
 \left[
 F_{h,0}+S_h/R_\gamma
 \right],
 &|D_0|<R_\gamma.
 \end{cases}
\end{equation*}
The parametric ROC is then
\begin{equation}
 p_{10}(\gamma;\bm r,t)
 =
 \int_{\cV_2}
 \mathcal T_0\dd^2\bm s,
 \qquad
 p_{11}(\gamma;\bm r,t)
 =
 \int_{\cV_2}
 \mathcal T_1\dd^2\bm s.
 \label{eq:qubit-parametric-roc}
\end{equation}
Equations~\eqref{eq:qubit-np-block-eigs}--\eqref{eq:qubit-parametric-roc}
retain both the continuous GKP syndrome and the noncommuting logical labels. For an axial displacement, a computational or Fourier state is sufficient for every $\gamma$, and the single-mode ROC equals the entanglement-assisted ROC. For an oblique displacement, the two statistical experiments generally differ.

\subsection{Finite-energy mixed GKP outputs}

For finite-energy states, use the quadrature nodes and orbit states from
Eq.~\eqref{eq:finite-orbit-kernel} and form the Gram matrix
\begin{equation*}
 G_{(h,r),(k,s)}
 =
 \sqrt{w_{hr}w_{ks}}\,
 \langle
 \Psi_{hr}^{(\tau)}
 |
 \Psi_{ks}^{(\tau)}
 \rangle.
\end{equation*}
Define
\begin{equation*}
 J_\gamma
 =
 \operatorname{diag}
 (-\gamma\id_{N_0},+\id_{N_1}),
 \qquad
 B_\gamma
 =
 G^{1/2}J_\gamma G^{1/2}.
\end{equation*}
The nonzero spectrum of
$\rho_1(t)-\gamma\rho_0$
is the spectrum of $B_\gamma$. Hence
\begin{equation*}
 \mathcal L_\omega(\gamma,t)
 =
 \Tr\pos{\rho_1(t)-\gamma\rho_0}
 =
 \sum_n
 \pos{\lambda_n[B_\gamma(t)]},
\end{equation*}
and
\begin{equation}
 P_{{\rm D},\omega}^{\star}(\alpha;t)
 =
 \min_{\gamma\geq0}
 \left[
 \gamma\alpha
 +\sum_n
 \pos{\lambda_n[B_\gamma(t)]}
 \right].
 \label{eq:finite-np-dual-gram}
\end{equation}
At a differentiable interior optimum,
\begin{equation*}
 -\partial_\gamma
 \mathcal L_\omega(\gamma,t)
 =p_{10}=\alpha,
 \qquad
 p_{11}
 =
 \mathcal L_\omega(\gamma,t)
 +\gamma\alpha.
\end{equation*}
The finite-energy mixed-state behavior is illustrated in
Fig.~\ref{fig:finite-np-tmin}(a). For the oblique displacement
considered there, finite squeezing lowers the detection probability
relative to the ideal GKP limit, but the entanglement-assisted protocol
continues to outperform the optimized single-mode GKP protocol over
the relevant false-alarm range. Increasing the squeezing progressively
restores the ideal syndrome--logical discrimination structure.

\subsection{Minimum detectable perturbation}

For a probe class $\mathcal P$, define
\begin{equation*}
 t_{\rm m}^{\mathcal P}
 (\bm u;\alpha,\zeta)
 =
 \inf
 \left\{
 t\geq0:
 \sup_{\omega\in\mathcal P}
 P_{{\rm D},\omega}^{\star}(\alpha;t)
 \geq\zeta
 \right\},
\end{equation*}
using the first crossing from the origin. A later crossing following a stabilizer revival represents a distinct detection window rather than a smaller local perturbation.

For pure probes, the threshold follows from
$\mathcal A_\omega(t)\geq a_\zeta(\alpha)$.
For the ideal entanglement-assisted protocol under compensated loss, one chooses the likelihood threshold in
Eqs.~\eqref{eq:bell-p10-param} and
\eqref{eq:bell-p11-param}
to satisfy
$p_{10}=\alpha$ and then solves for
$p_{11}=\zeta$, using
Eq.~\eqref{eq:theta-likelihood-ratio}.
Equation~\eqref{eq:unwrapped-gaussian-tmin}
provides the pre-alias Gaussian benchmark, while
Eq.~\eqref{eq:lmp-tmin}
gives the local threshold near the null. The optimized one-mode GKP-qubit threshold follows from
Eq.~\eqref{eq:qubit-parametric-roc}, and the finite-energy threshold from
Eq.~\eqref{eq:finite-np-dual-gram}.

\begin{figure*}[t]
\centering

\begin{minipage}[t]{0.49\textwidth}
    \centering
    \includegraphics[
        width=\linewidth
    ]{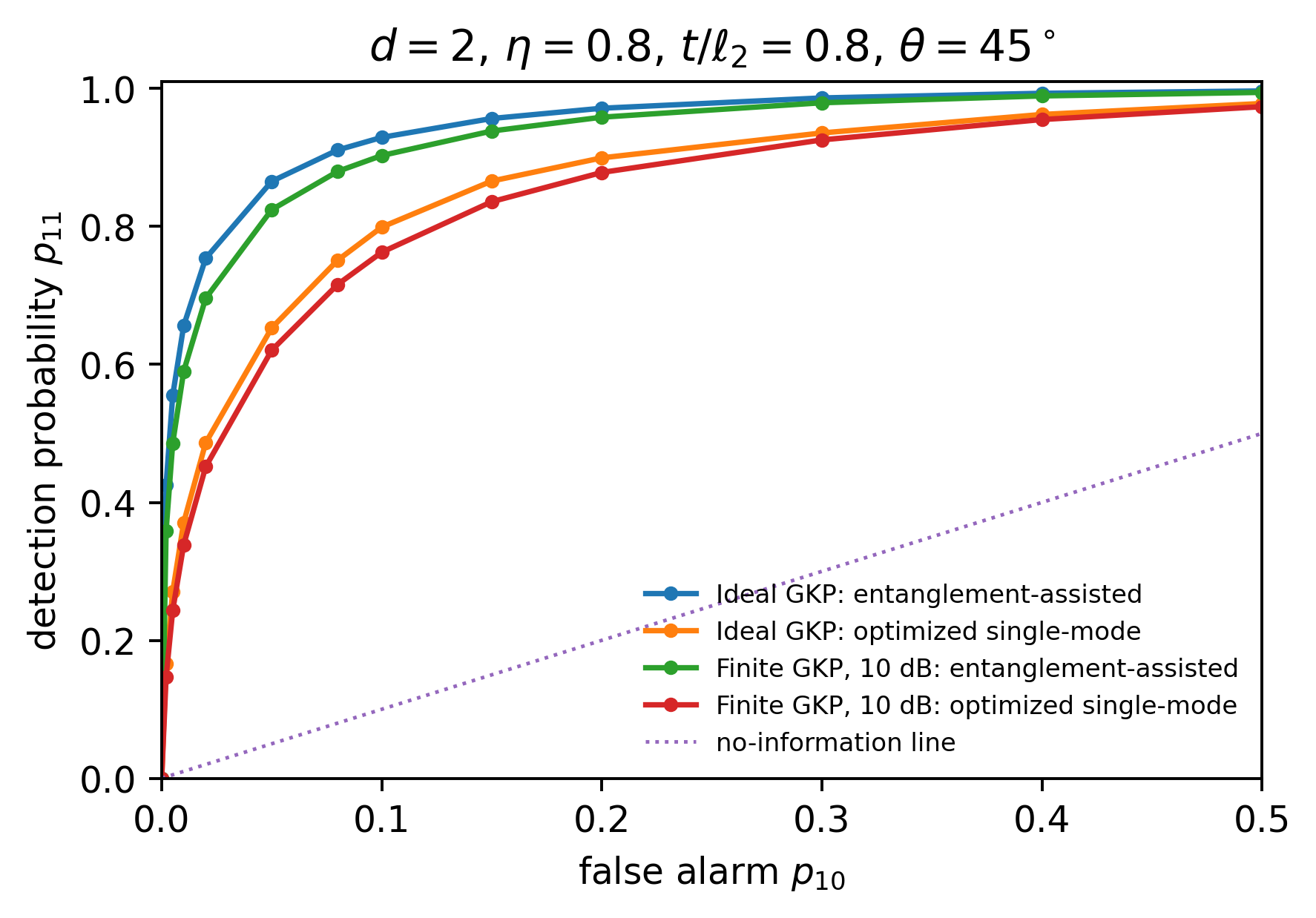}
    \vspace{-2mm}
    
    \textbf{(a)}
\end{minipage}
\hfill
\begin{minipage}[t]{0.49\textwidth}
    \centering
    \includegraphics[
        width=\linewidth
    ]{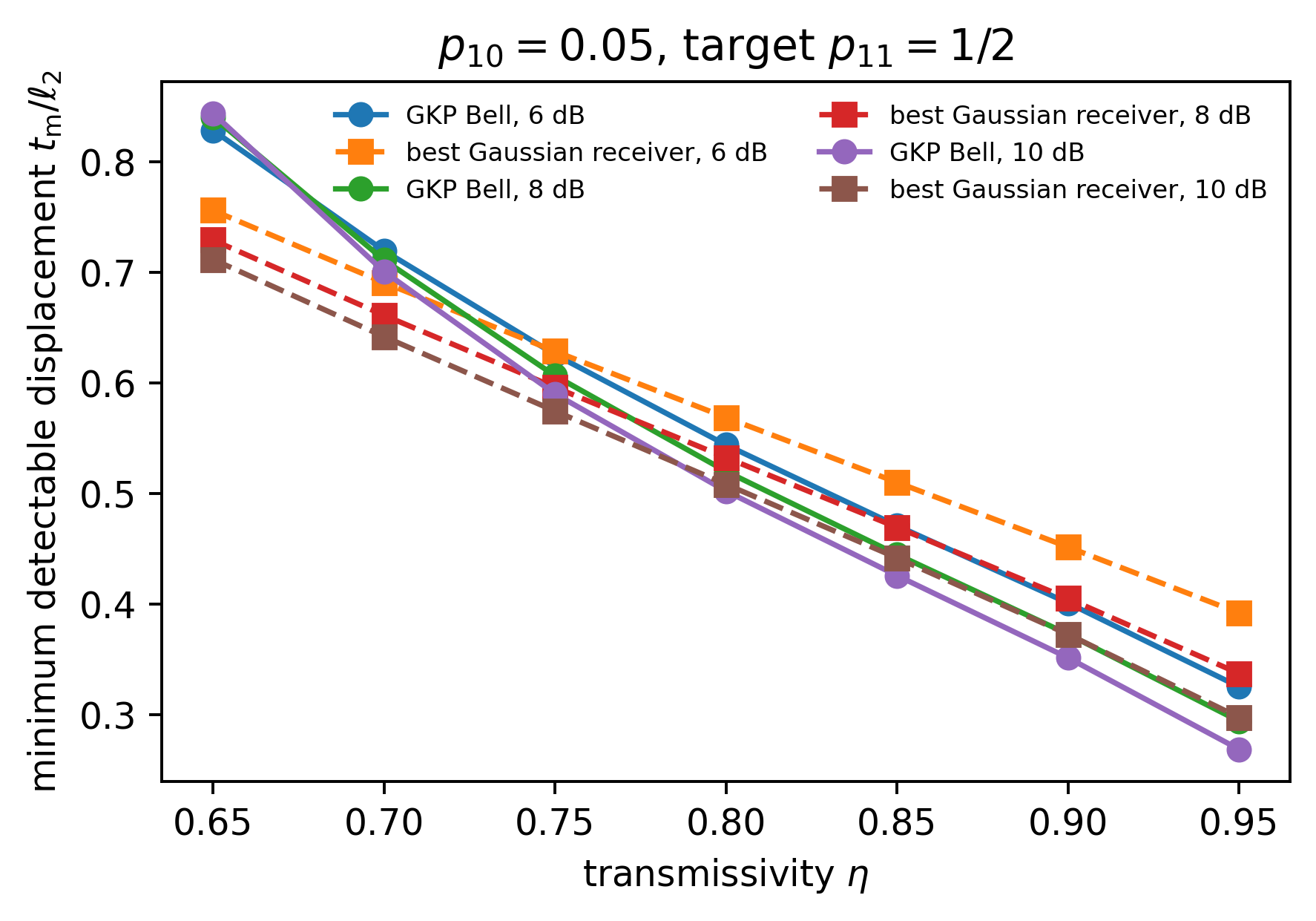}
    \vspace{-2mm}
    
    \textbf{(b)}
\end{minipage}

\caption{
Finite-energy Neyman--Pearson performance under post-amplified loss.
(a) Receiver-operating characteristics for an oblique displacement with
$d=2$, $\eta=0.8$, and $t/\ell_2=0.8$, comparing the ideal GKP limit
with the finite-energy entanglement-assisted and optimized single-mode
GKP protocols. Finite squeezing degrades the discrimination relative
to the ideal-code limit while preserving the separation between the
two protocols.
(b) First-crossing minimum detectable displacement as a function of
channel transmissivity for the finite-energy entanglement-assisted GKP
protocol at equal nominal squeezing. The false-alarm probability is
$p_{10}=0.05$ and the target detection probability is $p_{11}=1/2$.
Dashed curves denote the best selected Gaussian receiver at each
squeezing level. The crossing of the GKP and Gaussian curves identifies
the loss regime in which the GKP protocol provides the smaller
detectable perturbation.
}
\label{fig:finite-np-tmin}

\end{figure*}

The effect of loss on the first detectable perturbation is shown in
Fig.~\ref{fig:finite-np-tmin}(b). At fixed false-alarm and
target-detection probabilities, decreasing the transmissivity increases
the threshold for both the GKP and Gaussian protocols. For sufficiently
high transmissivity, the finite-energy GKP protocol achieves a smaller
minimum detectable displacement than the best selected Gaussian
receiver, whereas stronger loss eventually removes this advantage.
The crossover depends on the finite squeezing of the probe.

These results distinguish the two roles of GKP entanglement. It does not improve upon the pointwise optimized one-mode probe in the noiseless pure-state problem, but it removes preparation-dependent blind directions and preserves both logical displacement labels. Under loss and amplification, this additional record can improve the ROC and lower the minimum detectable perturbation for oblique displacements.

% ================================================================
\section{Discussion and conclusion}
\label{sec:conclusion}

We have developed a quantum-decision-theoretic framework for binary
phase-space displacement detection using finite-energy,
$d$-level GKP states. The framework treats both Bayesian
minimum-error discrimination and Neyman--Pearson hypothesis testing,
including the complete receiver-operating characteristic and the
first-crossing minimum detectable displacement. Finite squeezing is
incorporated through exact theta-series displacement kernels, rather
than by replacing the GKP states with ideal codewords or with a
purely classical modular-noise model. Pure loss followed by
quantum-limited amplification is mapped to a unit-gain additive
Gaussian channel, allowing the effects of attenuation, amplification
noise, logical aliases, and continuous GKP syndromes to be described
within a common statistical model.

The comparison between single-mode and entanglement-assisted probes
clarifies the role of GKP entanglement in displacement detection. In
the noiseless pure-state problem, entanglement does not outperform a
single-mode preparation optimized independently for each known
displacement. Its advantage is instead structural: a maximally
entangled GKP probe removes preparation-dependent blind directions
and converts the two noncommuting logical displacement labels into
orthogonal logical-Bell sectors. For axial displacements, an
appropriately chosen computational or Fourier single-mode state
retains all relevant information, and the two architectures coincide.
For oblique displacements, however, both logical labels generally
carry hypothesis information. Under finite squeezing, loss, and
amplification, preserving this joint record can improve the
receiver-operating characteristic and reduce the minimum detectable
perturbation relative to the optimized single-mode GKP strategy.

The numerical comparisons also identify regimes in which the
finite-energy GKP structure provides an advantage over the selected
Gaussian protocols. The advantage is not generally local around the
null hypothesis. A direction-matched squeezed state remains highly
effective for very small displacements, whereas the GKP improvement
appears near finite-displacement features associated with the modular
lattice structure. At $8$ dB, the largest Bayesian error reduction in
the investigated noisy parameter range is $0.03809$, obtained for
$d=5$ and $\eta=0.95$. In the Neyman--Pearson comparison at
$d=5$, $6$ dB, and $\eta=0.8$, the GKP protocol reduces the
first-crossing detectable displacement by $8.54\%$ relative to the
best of the selected Gaussian receivers. Increasing loss eventually
removes this advantage, demonstrating that the useful GKP regime is
set jointly by the finite squeezing, logical dimension, displacement
scale, and channel transmissivity.

These comparisons are made at equal nominal squeezing and against a
specified set of coherent-state, squeezed-vacuum, and twin-beam
receivers. An important next step is to optimize
the lattice geometry, logical dimension, probe preparation, and
receiver under equal signal-energy or total-energy constraints.
Further extensions include composite hypotheses with unknown
displacement direction, detector inefficiency and mode mismatch,
asymmetric and correlated displacement noise, and experimentally
accessible approximations to the optimal modular-syndrome and
logical-Bell measurements. More broadly, the results show that the
continuous syndrome and discrete logical structure of finite-energy
GKP states form complementary statistical resources, allowing
bosonic error-correcting codes to serve not only as protected
information carriers but also as structured probes for quantum
process detection.
% ================================================================
\appendix

\section{Finite-energy GKP theta-series matrix elements}
\label{app:theta}

\subsection{Direct Mehler-comb evaluation of the square-code kernel}

Equation~\eqref{eq:explicit-B-double-sum} follows directly from the finite-energy wavefunction in Eq.~\eqref{eq:finite-wavefunction}. For a pair of comb points, set
\begin{equation*}
 A=q_{jm},
 \qquad
 B=q_{kn}.
\end{equation*}
The required Gaussian integral is
\begin{align*}
 I_{AB}(x,p)
 &=
 \int_{\mathbb R}\dd q\,
 e^{-(q-c_\beta A)^2/(2v_\beta)}
 e^{-(q-x-c_\beta B)^2/(2v_\beta)}
 e^{ip(q-x/2)}
 \\
 &=
 \sqrt{\pi v_\beta}\,
 e^{-[c_\beta(A-B)-x]^2/(4v_\beta)}
 e^{-v_\beta p^2/4}
 e^{ic_\beta p(A+B)/2}.
\end{align*}
Multiplying by the envelope factors
\begin{equation*}
 e^{-v_\beta(A^2+B^2)/2}
\end{equation*}
and summing over $m,n\in\mathbb Z$ yields
Eq.~\eqref{eq:explicit-B-double-sum}. This representation is particularly convenient for numerical calculations with square GKP codes.

\subsection{Coset theta-series representation}

To expose the lattice and phase-sector structure, introduce normalized phase-space coordinates
\begin{equation*}
 \bm x=\frac{\bm\xi}{\sqrt{2\pi}},
 \qquad
 \mathsf D(\bm x)=D(\sqrt{2\pi}\bm x).
\end{equation*}
The Weyl relation becomes
\begin{equation}
 \mathsf D(\bm x)\mathsf D(\bm y)
 =
 e^{-i\pi\bm x^T\Omega\bm y}
 \mathsf D(\bm x+\bm y).
 \label{eq:norm-weyl}
\end{equation}

The ideal GKP projector is the distribution
\begin{equation}
 \Pi_\Lambda^\infty
 =
 \sum_{\bm\lambda\in\Lambda}
 e^{i\phi_M(\bm\lambda)}
 \mathsf D(\bm\lambda).
 \label{eq:ideal-projector}
\end{equation}
Within the ideal logical space, a matrix unit can be expanded in the logical Weyl basis as
\begin{equation}
 |k\rangle\langle j|
 =
 \frac{1}{d}
 \sum_{b=0}^{d-1}
 e^{-2\pi ibj/d}
 \bar X^{k-j}\bar Z^b.
 \label{eq:matrix-unit}
\end{equation}
Combining Eqs.~\eqref{eq:ideal-projector} and
\eqref{eq:matrix-unit} expresses each ideal logical matrix unit as a sum over logical cosets of the stabilizer lattice.

The Fock envelope has the Gaussian displacement representation \cite{ConradThesis2024}
\begin{equation*}
 \begin{aligned}
 e^{-\beta\hat n}
 &=
 C_\beta'
 \int_{\mathbb R^2}\dd^2\bm u\,
 e^{-\pi\norm{\bm u}^2/\Delta_\beta^2}
 \mathsf D(\bm u),
 \\
 \Delta_\beta^2
 &=2\tanh(\beta/2).
 \end{aligned}
\end{equation*}
Using Eq.~\eqref{eq:norm-weyl} twice, together with
$\Tr\mathsf D(\bm u)=\delta(\bm u)$, gives
\begin{equation*}
 \begin{split}
 &\Tr\!\left[
 \mathsf D^\dagger(\bm x)
 e^{-\beta\hat n}
 \mathsf D(\bm\lambda)
 e^{-\beta\hat n}
 \right]
 \\
 &\quad=
 C_\beta
 \exp\!\left[
 -\frac{\pi}{2\Delta_\beta^2}
 \norm{\bm x-\bm\lambda}^2
 -\frac{\pi\Delta_\beta^2}{8}
 \norm{\bm x+\bm\lambda}^2
 \right].
 \end{split}
\end{equation*}
Returning to physical phase-space coordinates identifies the Gaussian coefficients
\begin{equation*}
 a_\beta=\frac{1}{4\Delta_\beta^2},
 \qquad
 b_\beta=\frac{\Delta_\beta^2}{16}.
\end{equation*}

Let
\begin{equation*}
 e^{i\chi_{ab}}D(\bm c_{ab})
\end{equation*}
be the chosen physical representative of the logical Weyl operator $W_{ab}$. The corresponding lattice phase is
\begin{equation*}
 \Phi_{ab}(\bm\lambda)
 =
 \chi_{ab}
 +\phi_M(\bm\lambda)
 -\frac12
 \bm c_{ab}^T\Omega\bm\lambda.
\end{equation*}
Each finite-energy matrix element is therefore organized as a finite sum over logical characteristics and a Gaussian sum over the stabilizer lattice, i.e., a shifted lattice theta series. This form extends the direct square-code expression to a general symplectic lattice while retaining the projective stabilizer phases.

\section{Logical-basis derivation of the two-GKP Bell expansion}
\label{app:bell-expansion}

Let
\begin{equation*}
 H=G^{-1/2},
 \qquad
 |\bar j_\beta\rangle
 =
 \sum_k H_{kj}|w_{k,\beta}\rangle.
\end{equation*}
Substitution into Eq.~\eqref{eq:finite-bell-state-prelim} gives
\begin{equation*}
 |\Phi_{d,\beta}\rangle
 =
 \frac1{\sqrt d}
 \sum_{j,k,l}
 H_{kj}H_{lj}
 |w_{k,\beta}\rangle_A
 |w_{l,\beta}\rangle_R.
\end{equation*}
Its position-space wavefunction is
\begin{equation*}
 \Psi_{\Phi,\beta}(q_A,q_R)
 =
 \frac1{\sqrt d}
 \sum_{j,k,l}
 H_{kj}H_{lj}
 \psi_{k,\beta}(q_A)
 \psi_{l,\beta}(q_R).
\end{equation*}
Inserting Eq.~\eqref{eq:finite-wavefunction} for each factor gives the explicit finite-energy double-comb representation.

The maximally entangled-state identity
\begin{equation*}
 \langle\Phi_d|
 (A\otimes\id)
 |\Phi_d\rangle
 =
 \frac1d\Tr A
\end{equation*}
implies the Bell-sector coefficients in
Eq.~\eqref{eq:bell-expansion-coefficients}. Indeed,
\begin{align*}
 &\langle\Phi_{ab}^{(\beta)}|
 \bigl(
 P_AD_A(\bm\xi)P_A\otimes P_R
 \bigr)
 |\Phi_{d,\beta}\rangle
 \\
 &\qquad=
 \frac1d
 \Tr\!\left[
 W_{ab}^\dagger
 K_\beta^{(d)}(\bm\xi)
 \right].
\end{align*}
Since the logical Weyl operators form an orthogonal operator basis,
\begin{equation*}
 K_\beta^{(d)}(\bm\xi)
 =
 \sum_{a,b}
 c_{ab}(\bm\xi)W_{ab},
 \qquad
 \Tr(W_{ab}^\dagger W_{a'b'})
 =
 d\,\delta_{aa'}\delta_{bb'},
\end{equation*}
which yields Eq.~\eqref{eq:bell-in-code-weight}.

At an ideal logical displacement, only one Bell coefficient survives, giving Eq.~\eqref{eq:ideal-bell-label-explicit}. The Bell states are orthonormal because
\begin{align*}
 \langle\Phi_{ab}|\Phi_{a'b'}\rangle
 &=
 \frac{\delta_{aa'}}{d}
 \sum_{j=0}^{d-1}
 e^{2\pi i(b'-b)j/d}
 \\
 &=
 \delta_{aa'}\delta_{bb'}.
\end{align*}

\section{Loss and amplifier derivations}
\label{app:channels}

Applying Eqs.~\eqref{eq:loss-chi} and
\eqref{eq:amp-chi} sequentially with $G=1/\eta$ gives
\begin{equation*}
 \chi_{(\cA_{1/\eta}\circ\cL_\eta)(\rho)}(\bm k)
 =
 \chi_\rho(\bm k)
 e^{-(1-\eta)\norm{\bm k}^2/(2\eta)}.
\end{equation*}
This is the additive-noise channel in
Eq.~\eqref{eq:additive-channel}, with variance
\begin{equation*}
 \sigma_{\rm post}^2=\frac{1-\eta}{\eta}.
\end{equation*}
Reversing the order gives
\begin{equation*}
 \sigma_{\rm pre}^2=1-\eta.
\end{equation*}

For direct Fock-space propagation, the pure-loss Kraus operators are
\begin{equation*}
 L_\ell^{(\eta)}
 =
 \sum_{n=\ell}^{\infty}
 \sqrt{\binom n\ell}\,
 (1-\eta)^{\ell/2}
 \eta^{(n-\ell)/2}
 |n-\ell\rangle\langle n|,
\end{equation*}
while the quantum-limited amplifier Kraus operators are
\begin{equation*}
 A_m^{(G)}
 =
 \sum_{n=0}^{\infty}
 \sqrt{\binom{n+m}{m}}\,
 \frac{(G-1)^{m/2}}{G^{(n+m+1)/2}}
 |n+m\rangle\langle n|.
\end{equation*}
The post-amplified channel therefore acts as
\begin{equation*}
 \cA_{1/\eta}\circ\cL_\eta(O)
 =
 \sum_{m,\ell}
 A_m^{(1/\eta)}
 L_\ell^{(\eta)}
 O
 L_\ell^{(\eta)\dagger}
 A_m^{(1/\eta)\dagger}.
\end{equation*}

The no-loss Kraus operator is itself a GKP-type Fock envelope:
\begin{equation*}
 L_0^{(\eta)}
 =
 \eta^{\hat n/2}
 =
 e^{-\beta_\eta\hat n},
 \qquad
 \beta_\eta=-\frac12\ln\eta.
\end{equation*}
The higher-loss Kraus operators contain powers of $\hat a$ and generate the non-Pauli deformation absent from a classical random-displacement model.

\section{Gaussian inverse-preparation vacuum-or-not receivers}
\label{app:gaussian-nulling}

Equations~\eqref{eq:nulling-off-probability} and
\eqref{eq:nulling-np-roc} can be evaluated directly in phase space, without a Fock-space cutoff. We use a quadrature basis aligned with the known displacement direction $\bm u$. Exact nulling makes the transformed mean vanish under $H_0$ for all three probes.

For the coherent-state benchmark,
\begin{equation*}
 \widetilde V_{\rm coh}
 =
 \left(
 \frac12+\sigma^2
 \right)\id_2,
 \qquad
 \widetilde{\bm d}_{{\rm coh},1}
 =
 t\bm u.
\end{equation*}

For a squeezed vacuum with squeezed variance $v_s/2$ along $\bm u$, inverse squeezing gives
\begin{equation*}
 \widetilde V_{\rm sq}
 =
 \begin{pmatrix}
 \frac12+\sigma^2e^{2r} & 0\\
 0 & \frac12+\sigma^2e^{-2r}
 \end{pmatrix},
 \qquad
 \widetilde{\bm d}_{{\rm sq},1}
 =
 \begin{pmatrix}
 e^rt\\
 0
 \end{pmatrix}.
\end{equation*}

For the twin-beam state, define
\begin{equation*}
 c_r=\cosh r,
 \qquad
 s_r=\sinh r.
\end{equation*}
Undoing the two-mode squeezing transformation gives
\begin{align*}
 \widetilde V_{\rm TWB}
 &=
 \frac12\id_4
 +
 \sigma^2
 \begin{pmatrix}
 c_r^2\id_2 & -c_rs_rZ\\
 -c_rs_rZ & s_r^2\id_2
 \end{pmatrix},
 \\
 \widetilde{\bm d}_{{\rm TWB},1}
 &=
 t
 \begin{pmatrix}
 c_r\bm u\\
 -s_rZ\bm u
 \end{pmatrix}.
\end{align*}

For an $m$-mode Gaussian state with covariance matrix $V$ and mean $\bm d$, the vacuum probability is
\begin{equation*}
 \langle 0^{\otimes m}|
 \rho
 |0^{\otimes m}\rangle
 =
 \frac{
 \exp\!\left[
 -\bm d^T
 (V+\id/2)^{-1}
 \bm d/2
 \right]
 }{
 \sqrt{\det(V+\id/2)}
 }.
\end{equation*}
Substitution of the covariances and means above yields
Eq.~\eqref{eq:nulling-off-probability}. Although the preparation and inverse transformation are Gaussian, the complete receiver is non-Gaussian because vacuum-or-not detection is not a Gaussian measurement.

\section{Poisson summation and wrapped GKP likelihoods}
\label{app:poisson}

For a lattice $\Lambda$ and its symplectic dual
$\Lambda^\perp$, Poisson summation gives \cite{ConradThesis2024}
\begin{equation}
 \sum_{\bm\lambda\in\Lambda}
 f(\bm\lambda)
 =
 \frac{1}{\operatorname{vol}(\Lambda)}
 \sum_{\bm\lambda^\perp\in\Lambda^\perp}
 \widehat f(\bm\lambda^\perp),
 \label{eq:poisson}
\end{equation}
where $\widehat f$ denotes the symplectic Fourier transform.

Applying Eq.~\eqref{eq:poisson} to the Gaussian images in
Eq.~\eqref{eq:wrapped-density-bayes} gives a rapidly convergent Fourier representation. In one dimension,
\begin{align*}
 f_{h,q}^{a}(s_q)
 &=
 \sum_{m\in\mathbb Z}
 \frac{
 e^{-[s_q+\ell_d(a+dm)-\xi_{h,q}]^2/(2\sigma^2)}
 }{
 \sqrt{2\pi}\sigma
 }
 \\
 &=
 \frac{1}{d\ell_d}
 \sum_{r\in\mathbb Z}
 e^{-2\pi^2\sigma^2r^2/(d^2\ell_d^2)}
 \\
 &\qquad\times
 e^{2\pi ir[s_q+\ell_da-\xi_{h,q}]/(d\ell_d)}.
\end{align*}
Recognizing this Fourier series as a Jacobi theta function gives the wrapped-density expression used in
Eq.~\eqref{eq:wrapped-theta-factor}.

For isotropic noise, the two-dimensional likelihood factorizes:
\begin{equation*}
 f_h^{ab}(\bm s)
 =
 f_{h,q}^{a}(s_q)
 f_{h,p}^{b}(s_p).
\end{equation*}
An axial displacement changes only one factor, which establishes the single-mode sufficiency and axial-equality results used in the Bayesian and Neyman--Pearson analyses.

\section*{Data availability}

The numerical data supporting the findings of this study are
available from the corresponding author upon request.

\section*{Author contributions}

S.K. conceived the study, developed the theoretical framework,
performed the numerical analysis, prepared the figures, and wrote
the original manuscript. S.C. supervised the work, contributed to
the interpretation of the results, and reviewed and edited the
manuscript. Both authors read and approved the final manuscript.

\section*{Competing interests}

The authors declare no competing interests.
\bibliography{references}

\end{document}